\documentclass[galaxies,review,accept,pdftex,moreauthors]{Definitions/mdpi}
\firstpage{1} 
\pubvolume{1}
\issuenum{1}
\articlenumber{0}
\pubyear{2026}
\copyrightyear{2026}
\externaleditor{Firstname Lastname} 
\datereceived{7 April 2026} 
\daterevised{17 May 2026} 
\dateaccepted{18 May 2026} 
\datepublished{ } 
\makeatletter
\renewcommand{\linenumbers}{}
\makeatother

\Title{A Review on Resolving the Hubble Tension via \mbox{Late-Universe Physics}}

\Author{Xuan-Dong Jia $^{1}$\orcidJ{}, 
 Xin-Yi Dai $^2$, Yu-Peng Yang $^{2}$\orcidY{} and Fa-Yin Wang $^{1,}$*\orcidW}

\AuthorNames{Xuan-Dong Jia, Xin-Yi Dai,  Yu-Peng Yang and Fa-Yin Wang}

\address{%
$^1$\quad School of Astronomy and Space Science, Nanjing University, Nanjing 210023, China; 602023260009@smail.nju.edu.cn 
\\
$^{2}$ \quad School of Physics and Physical Engineering, Qufu Normal University, Qufu 273165, China; 1584377516@qq.com (X.-Y.D.); 
 ypyang@aliyun.com (Y.-P.Y.)}

\corres{Correspondence: fayinwang@nju.edu.cn}

\abstract{The $\Lambda$CDM cosmological model has been successful in explaining many astronomical observations. However, recent observations increasingly point to deviations from the standard 
$\Lambda$CDM framework. Among these, one of the most significant discrepancies is the \textit{Hubble 
tension}, which refers to the difference in values obtained 
for the Hubble constant $H_0$ from high-redshift measurement and local observation. To address this issue, numerous cosmological models and methodological approaches have been proposed. This review offers a concise overview of recent progress in resolving the Hubble tension. The combination of Dark Energy Spectroscopic Instrument (DESI) Baryon Acoustic Oscillations (BAO) and uncalibrated Type Ia supernovae data yields a value for $H_0$ that is significantly higher than the $\Lambda$CDM predication based on early-universe probes, even without incorporating local distance ladder constraints. This result indicates that the origin of the Hubble tension lies in new physics at low redshifts. Our findings suggest that although many unresolved systematics persist in current observations, they are insufficient to account for the magnitude of the current Hubble tension. This implies the likely 
existence of new physical mechanisms that have yet to be discovered.
}

\begin{document}

\newpage
\tableofcontents
\newpage

\section{Introduction}
The standard cosmological model---known as $\Lambda$CDM---combines a cosmological constant ($\Lambda$) with cold dark matter (CDM). It 
has achieved remarkable success in describing a wide array of cosmological observations, most notably the exquisitely precise measurements of the cosmic microwave background (CMB). Within the $\Lambda$CDM framework, the $Planck$ collaboration infers a Hubble constant of 
$H_0 = 67.4 \pm 0.5$ km s$^{-1}$ Mpc$^{-1}$ \cite{2020A&A...641A...6P}. In contrast, local determinations of $H_0$ adopt a largely model-independent approach, relying on Local Distance Ladder methods that do not assume a specific cosmological model. These yield a significantly higher value: $H_0 = 73.17 \pm 0.86$ km s$^{-1}$ Mpc$^{-1}$ \cite{2022ApJ...934L...7R,2024ApJ...973...30B}. The discrepancy between these two results amounts to approximately $5.8 \sigma$. While different local measurement techniques may produce slightly varying estimates of $H_0$, the tension remains statistically robust---persisting at or above the $\sim$$4 \sigma$ level even in the most conservative \mbox{comparisons \cite{2018ApJ...855..136R,2019ApJ...876...85R,2020A&A...641A...6P,2020NatRP...2...10R,2020MNRAS.498.1420W,2021MNRAS.502.2065D,2021ApJ...908L...6R,2022ApJ...934L...7R,2023ApJ...954L..31S,2024ApJ...977..120R,2024arXiv241000804B,2025ApJ...992L..34R,2025ApJ...988...97L,2025ApJ...979L...9S}.} This growing tension has become one of the most significant puzzles in modern cosmology, as increasingly precise measurements of $H_0$ over the past decade have revealed a striking and persistent discrepancy between its value inferred from observations of the early Universe and its value measured directly in the late Universe (for reviews, see \cite{2022JHEAp..34...49A,2022Univ....8..399D,2022NewAR..9501659P,2023JCAP...07..020K,2024ARA&A..62..287V}). This persistent inconsistency is widely known as the ``Hubble Tension''.

The Hubble constant, $H_0$, quantifies the current expansion rate of the Universe, thereby serving as a cornerstone for deriving key cosmological quantities. Inferring the Hubble constant from the CMB is fundamentally distinct from local measurement methods. Observations of the CMB probe the universe at a redshift of approximately $z \approx 1100$. The CMB is sensitive to the expansion rate of the universe at the epoch of recombination. Assuming the standard cosmological model, constraints derived from the recombination era can be consistently extrapolated forward to infer constraints on the present-day universe. This approach yields tight constraints on cosmological parameters \cite{2020A&A...641A...6P}. However, it is inherently model-dependent. If the cosmic expansion history between recombination and today deviates from the standard model, the inferred value of $H_0$ would consequently change. Local determinations of the Hubble constant $H_0$ rely primarily on the local distance ladder, a hierarchical framework in which distances are measured step by step using a sequence of calibrated standard candles \cite{2022ApJ...934L...7R,2023ApJ...954L..31S,2024ApJ...977..120R,2025ApJ...992L..34R,2025ApJ...988...97L}. Each rung of the ladder depends on the one below it: nearby distance indicators---such as those based on parallax---are used to calibrate the intrinsic luminosities of brighter standard candles (e.g., Cepheid \mbox{variables \cite{2022ApJ...934L...7R,2022ApJ...938...36R,2022ApJ...940...64Y,2024ApJ...973...30B}}) residing in the same galaxies. These, in turn, calibrate even more luminous objects (such as Type Ia supernovae, SNe Ia \cite{Pantheon+,DESY5,Union3}) that can be observed at cosmological distances. Because standard candles have well-understood properties, comparing these with their observed fluxes yields direct distance estimates. However, this approach is inherently cumulative: any systematic error---or potential new physics---affecting a single rung propagates through all subsequent steps, potentially biasing the final measurement of $H_0$. Consequently, the robustness of the local $H_0$ value hinges critically on the accuracy and calibration of every link in the distance ladder, except for one-step methods like Type II SNe \cite{2025A&A...702A..41V}. 

In addition to analyses of potential systematic errors in the local distance ladder, numerous cosmological models have been proposed in the late Universe to explain the Hubble tension. In the standard cosmological model, as the simplest possible framework, dark matter is treated as a cold, pressureless, collisionless fluid, and dark energy is represented by a cosmological constant \cite{2016IJMPD..2530007B,2022JHEAp..34...49A,2022Univ....8..399D,2022NewAR..9501659P,2023CQGra..40i4001K,2023JCAP...07..020K,2026arXiv260101525D}. Recent measurements from CMB \cite{2020A&A...641A...6P,2025arXiv250620707C,2025JCAP...11..062L}, baryon acoustic oscillation (BAO) \cite{2021PhRvD.103h3533A,DESI_DR1,DESI_DR2}, as well as SNe Ia compilations \cite{Pantheon+,DESY5,Union3}, all show that the $\Lambda$CDM model provides an excellent fit when each probe is considered individually. Despite this overall success, the $\Lambda$CDM model yields a value of the Hubble constant that is in significant tension with local distance ladder measurements. The Hubble constant is the present-day expansion rate of the Universe, inferred by combining observational data with the assumed cosmic expansion history. If we assume that the observational data are free of issues, then modifying the expansion history of the Universe would alter the inferred value of the Hubble constant. Di Valentino et al. \cite{2021CQGra..38o3001D} gave a detailed classification for \mbox{the solutions.} 

In this review, we summarize various approaches to resolving the Hubble tension by analyzing systematic uncertainties in late Universe observational methods as well as by modifying the cosmological model. This review is organized as follows. In Section \ref{Sec:Measurements and uncertainties of the Hubble constant}, we briefly present the values of $H_0$ obtained from various measurement methods. In Section \ref{Sec:Explanations for the Hubble tension}, we review different cosmological models proposed to explain the Hubble tension. Finally, in Section \ref{Sec:Summary}, we provide a discussion and summary.

\section{Measurements and Uncertainties of the Hubble Constant} \label{Sec:Measurements and uncertainties of the Hubble constant}
The Hubble constant represents the current expansion rate of the universe. Nearby systems are primarily governed by gravitational interactions, whereas distant galaxies are more sensitive to the expansion of the Universe. By measuring the distances and redshifts of astronomical objects, one can infer the Hubble constant. Among these, the measurement of the redshift is often relatively straightforward, whereas the estimation of distance is typically the more challenging aspect. For objects that are relatively close, distance determinations can rely on reliable methods like parallax. However, for more distant objects---beyond the resolution limits of today's instruments---a common and conceptually simple strategy is to infer distance by comparing what we observe with their intrinsic properties. This approach essentially shifts the challenge from measuring distance directly to estimating the object's inherent characteristics. Astronomical objects with known intrinsic brightness (standard candles) or known physical size (standard rulers) serve as crucial pillars in the measurement of the Hubble constant. Currently, measurements of $H_0$ are generally categorized into model-independent and model-dependent approaches. In the following sections, we will review various methods for determining $H_0$, proceeding from model-independent to model-dependent techniques.

\subsection{Local Distance Ladder}
The most effective current approach to measuring the Hubble constant is constructing a local distance ladder based on a sequence of distance indicators \cite{2022ApJ...934L...7R}. The local distance ladder is not a single probe, but rather a combination of different distance measurement techniques. By cross-calibrating probes at overlapping distance ranges, it enables increasingly precise and more distant measurements. The Hubble Space Telescope (HST) Key Project carried out the first such measurements and achieved a precision of $10\%$ by \mbox{2001 \cite{2001ApJ...553...47F,2006ApJ...653..843S}.} Since then, the results have been significantly improved due to the accumulation of observational samples, unified calibration procedures, and near-infrared observations of Cepheid variables \cite{2016ApJ...826...56R,2018ApJ...855..136R,2019ApJ...876...85R,2021ApJ...908L...6R}. The latest measurement yields \mbox{$H_0 = 73.17 \pm 0.86$ $\rm km s^{-1} Mpc^{-1}$,} representing the most precise local measurement of $H_0$ to date \cite{2022ApJ...934L...7R,2024ApJ...973...30B}. 

\begin{figure}[H]
    \includegraphics[width=0.9\textwidth]{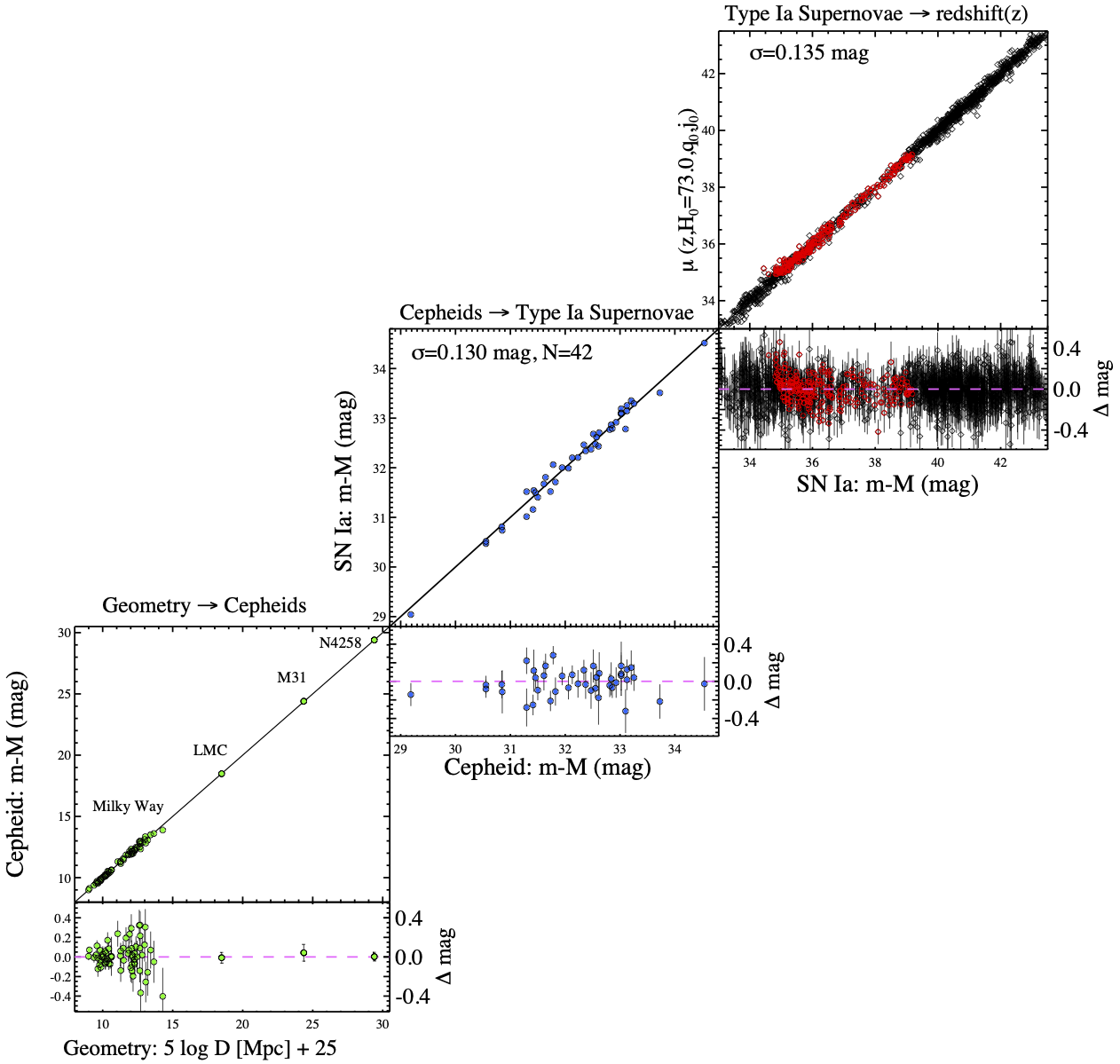}
    \caption{Three-
 rung distance ladder. The lower-left panel depicts geometric distance measurements to Cepheid variables, which are used to calibrate their period--luminosity relationship. In the central panel, SNe Ia in host galaxies that also contain Cepheids are calibrated against these Cepheid-derived distances. Finally, the upper-right panel displays SNe Ia residing in the Hubble flow; once the calibration is established, those meeting the specified redshift criteria provide a direct determination of the Hubble constant. Figure adapted from \cite{2022ApJ...934L...7R}.}\label{Fig:H0_Riess2022}
\end{figure}

While numerous methodologies exist for constructing a distance ladder to determine the Hubble constant, the most prevalent and precise framework relies on three rungs \cite{2022ApJ...934L...7R}. We will use this as a case study to introduce the distance ladder. The first rung of the distance ladder typically involves geometric distance measurements in the Milky Way (MW) \cite{2023A&A...674A...1G} or nearby galaxies---such as the Large Magellanic Cloud (LMC) \cite{2019Natur.567..200P}, Small Magellanic Cloud (SMC) \cite{2020ApJ...904...13G}, or NGC 4258 \cite{2019ApJ...886L..27R}. These galaxies are commonly referred to as ``anchors''. In the second rung, the period--luminosity relation of Cepheid variables in nearby galaxies is calibrated using the geometric distances established in the first rung. This calibrated Cepheid relation is subsequently used to determine distances to SNe Ia residing in the same host galaxies. Finally, in the third rung, the Hubble constant is measured directly from the distances and redshifts of SNe Ia located in the Hubble flow, where cosmic expansion dominates over local peculiar motions. These are illustrated in Figure \ref{Fig:H0_Riess2022}. 

\vspace{-5pt}

\begin{figure}[H]
    \includegraphics[width=0.9\textwidth]{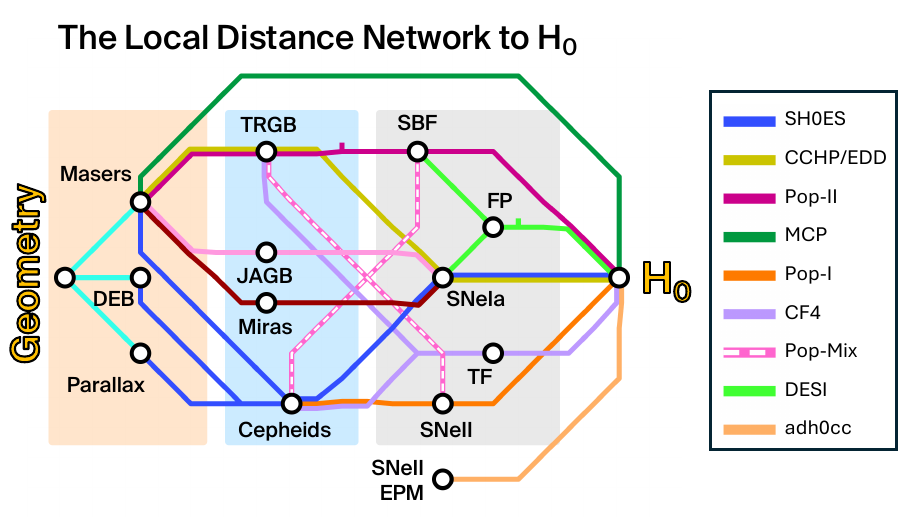}
    \vspace{-8pt}
    \caption{Conceptual 
 overview of the Local Distance Network. The Hubble constant is ultimately measured by combining different distance measurement methods. A non-exhaustive list of baseline linkages is labeled on the right. Background rectangles in orange, light blue, and gray indicate the first, second, and third rungs, respectively. Figure is adapted from \cite{2025arXiv251023823H}. For more detailed information, please refer to it. The references include SH0ES \cite{2022ApJ...934L...7R}, CCHP \cite{2025ApJ...985..203F}, EDD \cite{2021AJ....162...80A}, Pop-II \cite{2024ApJ...973...83A}, MCP \cite{2020ApJ...890..118P}, Pop-I \cite{2022MNRAS.514.4620D}, CF4 \cite{2020ApJ...902..145K}, DESI \cite{2025MNRAS.539.3627S}, adh0cc \cite{2025A&A...702A..41V}.} \label{Fig:local_distance_network}
\end{figure}

Although this method provides the tightest constraints to date, the formulation of the distance ladder is not unique. Cepheid variables in the first and second rungs of the distance ladder can be replaced by other standard candles, such as the Tip of the Red Giant Branch (TRGB) \cite{2019ApJ...882...34F,2024ApJ...966...89A}, the J-region magnitude of Asymptotic Giant Branch stars (JAGB) \cite{2024ApJ...961..132L,2024ApJ...966...20L}, and Mira variables \cite{2020ApJ...889....5H,2024ApJ...963...83H}. SNe Ia in the second and third rungs can also be replaced by the Surface Brightness Fluctuation (SBF) \cite{2021ApJ...911...65B,2024ApJ...973...83A}, Type II supernovae \cite{2022MNRAS.514.4620D,2023A&A...672A.129C}, and the Tully--Fisher relation \cite{2022MNRAS.511.6160K}. Different distance indicators can even be combined into a local distance network to measure the Hubble constant, shown in Figure \ref{Fig:local_distance_network} \cite{2025arXiv251023823H}.

\subsubsection{First Rung of the Ladder}
Galaxies within the first rung of the local distance ladder are typically referred to as ``anchors''. These systems, which include the MW and nearby galaxies hosting geometric measurements, establish the baseline for the distance ladder, thereby providing the fundamental calibration for determining the Hubble constant \cite{2023A&A...674A...1G,2019Natur.567..200P,2020ApJ...904...13G,2019ApJ...886L..27R}. Currently, the primary calibrators employed on this first rung mainly encompass the following four categories: $Gaia$ DR3 parallaxes of Cepheids and their host clusters in the MW \cite{2023A&A...674A...1G}; distances from detached eclipsing binaries in the LMC and SMC \cite{2019Natur.567..200P,2020ApJ...904...13G}; and the water maser-based distance to NGC 4258 \cite{2019ApJ...886L..27R}.

\vspace{-12pt}

Precise geometric distance measurements provide the basis for the standard candle calibration of Cepheids. Due to their well-defined period--luminosity (P-L) relation, Cepheids are widely regarded as excellent distance indicators \cite{1912HarCi.173....1L}. The relation is commonly expressed as: 
$ M = a \ \textrm{log} P + b$, where $M$ is the absolute magnitude, $P$ is the pulsation period, and $a$ and $b$ are constants. By measuring the pulsation period of a Cepheid variable, one can infer its absolute magnitude $M$. Combining this with an observation of its apparent magnitude $m$ then yields the distance $D$ via the distance modulus relation: $m-M = 5 \ \textrm{log}_{10}(D) + 25$, where $D$ is in Mpc.

Although current measurements of Cepheids are highly precise, they are still subject to systematic errors. Cepheids play a pivotal role in linking the first and second rungs of the distance ladder \cite{2022ApJ...934L...7R}. The P-L relation is initially calibrated locally via geometric distance measurements and subsequently employed to calibrate SNe Ia on the second rung of the distance ladder. However, systematic uncertainties in Cepheid distances may stem from disparities between nearby calibrators and the more distant Cepheids residing in SNe Ia host galaxies. These include differences in chemical composition, the range of pulsation periods, stellar crowding, dust properties, and potential nonlinearities in the P-L \mbox{relation \cite{2024arXiv240302801A}.} 

In observations of Cepheids, the systematic uncertainties encountered and the corresponding improvements proposed include: 

\textbf{1. Photometric systems.} 
Combining multiple photometric systems to observe Cepheids in the first and second rungs introduces a systematic uncertainty of $1.4$--$1.8\%$ 
in distance measurements \cite{2019ApJ...876...85R}. Observing Cepheids with the same telescope and instrument minimizes calibration uncertainties to the greatest extent possible. 

\textbf{2. Crowding.} Resolving Cepheids from their surrounding stellar populations is one of the major challenges in measuring Cepheids within SNe Ia host galaxies \cite{2019ApJ...882...34F}. 
With the advent of the James Webb Space Telescope (JWST), the ability to resolve Cepheids from background stars has been significantly enhanced. Recent JWST observations are consistent with previous HST measurements, reducing the dispersion of the P-L relation from $0.45$~to $0.17$ mag \cite{2023ApJ...956L..18R}. \citet{2024ApJ...962L..17R} indicate that JWST observations rule out unrecognized crowding in HST photometry that increases with distance as the cause of the Hubble tension, with a confidence level of $8.2 \sigma$.

\textbf{3. Metallicity.}
The luminosity of Cepheids at a given period is correlated with their metallicity, but this dependence is difficult to constrain \cite{1998ApJ...498..181K,2004ApJ...608...42S,2006ApJ...652.1133M}. Accounting for metallicity effects is essential in the distance ladder. Although Cepheids in NGC $4258$ and the 
MW closely resemble those in SNe Ia host galaxies---making metallicity differences negligible for determining $H_0$---the situation is different for the Magellanic Clouds, whose Cepheids are significantly more metal-poor than those in typical host galaxies \cite{2019ApJ...876...85R,2024ApJ...973...30B}. Since the metallicity of Cepheids in SNe Ia host galaxies closely matches those in the anchor galaxies, the Hubble constant is only minimally affected by metallicity effects. Neglecting this correction would alter $H_0$ by merely $0.5$ km s$^{-1}$ Mpc$^{-1}$ \cite{2022ApJ...934L...7R}.

\textbf{4. Dust laws.} Dust has long represented a major source of systematic uncertainty in the local distance ladder \cite{2018MNRAS.477.4534F}. To account for dust extinction, one can either apply a correction of $R_\lambda \times E(V-I)$ to the observed Cepheid magnitudes, or instead use a reddening-free Wesenheit magnitude \cite{1982ApJ...253..575M,2022ApJ...934L...7R}. Moreover, beyond the uncertainties inherent in the reddening term $E(V-I)$, the value of the extinction parameter $R_V$ is not independently measured for individual SNe Ia host galaxies or at each Cepheid location; instead, it is typically assumed to be the same as in the MW \cite{2021PhRvD.104l3511P,2022ApJ...934L...7R}. A complete sampling of the reddening curve at long and short wavelengths in each host galaxy, together with a more precise characterization of $R_V$, could further mitigate the uncertainty in Cepheid distance measurements \cite{2022ApJ...926..122H}.

\subsubsection{Second Rung of the Ladder}
The Cepheid P-L relation of the Cepheid is calibrated using the anchor galaxies in the first rung of the distance ladder. This calibration enables the conversion of Cepheid pulsation periods observed in distant SNe Ia host galaxies into distance measurements. These distances are then used to calibrate SNe Ia, which in turn serve as standardizable candles in the third rung for measuring the Hubble constant. In an ideal scenario, one could bypass the intermediate Cepheid step and directly link geometric calibrations to supernova-based distance measurements. Unfortunately, the rate of SNe in the local universe is far too low to provide enough SNe Ia events within the anchor galaxies.

SNe Ia are excellent standard candles, which originate in binary systems. In the canonical scenario, a carbon-rich white dwarf accretes matter from a companion until its mass approaches or exceeds the Chandrasekhar limit, triggering carbon fusion and a thermonuclear explosion. An alternative channel involves the merger of two white dwarfs whose combined mass is close to the Chandrasekhar mass. Because these explosions are powered by progenitors of nearly identical mass and involve similar physical mechanisms, the resulting light curves exhibit remarkable uniformity. Within the SALT modeling \mbox{framework \cite{2007A&A...466...11G,2021ApJ...923..265K},} the standardized apparent brightness of SNe Ia is typically derived using the Tripp formula \cite{1998A&A...331..815T} as follows: 
\begin{equation}
    m_X = m_B + \alpha x_1 - \beta c - \delta_{\text{Bias}} + \delta_{\text{Host}},
\end{equation}
where $m_B$, $x_1$, and $c$ are independent parameters derived from the SALT light-curve fitting model, $\alpha$ and $\beta$ denote the correlation coefficients, $\delta_{\text{Bias}}$ represents the correction for selection effects and other biases predicted by simulations, and $\delta_{\text{Host}}$ accounts for the correction due to host galaxy properties.

$H_0$ measurements derived from SNe Ia are subject to various sources of systematic uncertainty. The local distance ladder utilizes identical indicators across multiple rungs. Consequently, while systematic discrepancies between these rungs can severely bias the $H_0$ determination, uniform systematic shifts propagate coherently through the ladder and largely cancel out, leaving the final result virtually unaffected.

In the second rung of the distance ladder, systematic uncertainties affecting SNe \mbox{Ia include:} 

\textbf{1. Calibration.} In scenarios where a unified survey probes SNe in both the second and third distance ladder rungs, the aforementioned effect attenuates calibration systematics. However, this cancellation mechanism is absent when utilizing different surveys, thereby demanding more prudent consideration of systematic errors arising from calibration \cite{2018ApJ...869...56B,2019ApJ...882...34F}.

\textbf{2. SN Physics.} 
The intrinsic astrophysics of SNe and the impact of dust remain among the least constrained aspects of SN cosmology, potentially introducing systematic uncertainties into $H_0$ measurements \cite{2023JCAP...11..046M,2023ApJ...945...84P}. Most analyses indicate that these systematics have a minimal impact on $H_0$, as the discrepancies between the SNe in the second and third rungs of the distance ladder are not expected to be significant. For instance, \mbox{\citet{2015ApJ...802...20R}} indicated that there is a correlation between the standardized brightness of SNe and the age of their host galaxies. Additionally, discrepancies in dust extinction or color may arise between SNe in the second and third rungs \cite{2024MNRAS.533.2319W}. Employing SNe data with similar properties helps mitigate these systematics.

\begin{figure}[H]
    \hspace{-1mm}\includegraphics[width=0.8\textwidth]{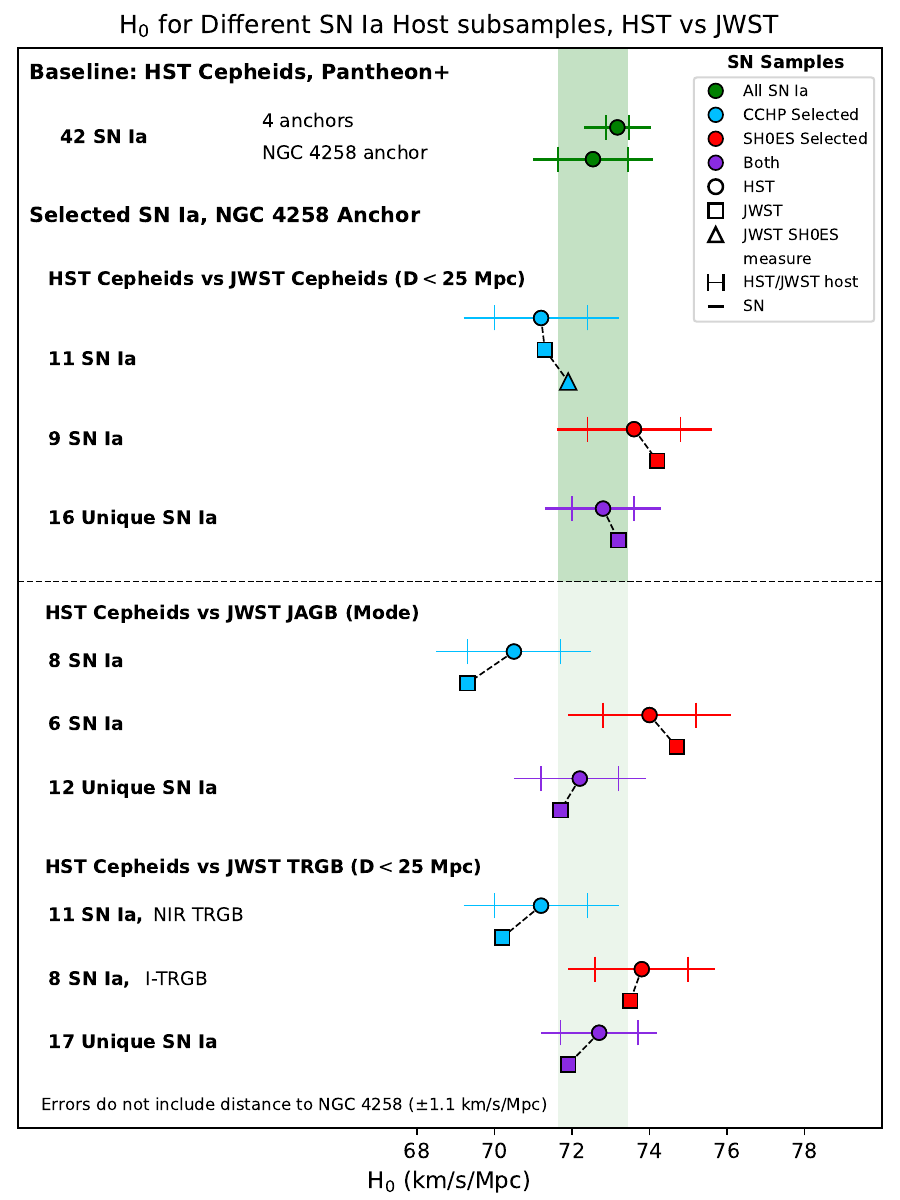}
    \vspace{-3pt}
    \caption{Comparisons 
 of $H_0$ between \textit{HST} 
Cepheids and other measures (\textit{JWST} Cepheids, \textit{JWST} JAGB, and \textit{JWST} NIR-TRGB) for SN~Ia host subsamples. The data demonstrate that \textit{HST} and \textit{JWST} distance measurements are in good agreement, with the selection of different subsamples by teams (CCHP vs. SH0ES) accounting for the primary variations rather than instrumental systematics or differences in distances to galaxies in common. Figure adapted from Riess et al. (2024)~\cite{2024ApJ...977..120R}.}
    \label{fig:hst_jwst_h0}
\end{figure}

\textbf{3. Selection effect.} The selection of the SNe sample may introduce a bias in the inferred value of $H_0$. Since the number of SNe Ia currently used for calibration in the second rung is still limited, selecting only a subset of these SNe may introduce a bias, potentially shifting the final inferred value of $H_0$. The results obtained using different telescopes and SNe Ia samples are shown in Figure \ref{fig:hst_jwst_h0}.

\subsubsection{Third Rung of the Ladder}
Type Ia supernovae are sufficiently abundant that, once the relationship between their intrinsic luminosities and observed light curves is calibrated—enabling robust distance estimates—they can be used in the third rung of the cosmic distance ladder to precisely trace the distance–redshift relation in the Hubble flow. In this regime, typically spanning distances of 100–600 Mpc (corresponding to redshifts 
$z < 0.15$), their calibrated brightnesses anchor the Hubble–Lemaître law, where the smooth expansion of the universe dominates over local peculiar motions of galaxies. 

In the third rung of the ladder, the SNe observed beyond the calibrator host galaxies are used to measure luminosity distances deep into the Hubble flow via
\begin{equation}
    D_L(z) = \frac{1}{H_0}(1+z)\int_0^z \frac{\mathrm{d}z'}{E(z')},
\end{equation}
where $E(z) \equiv H(z)/H_0$. Over the relatively low-redshift range up to $z < 0.15$, the distance-redshift relation can be accurately described as
\begin{equation}\label{eq:3}
    \log D_L(z) \approx \log z \left[ 1 + \frac{1}{2}(1 - q_0)z - \frac{1}{6}(1 - q_0 - 3q_0^2 + j_0)z^2 \right] - \log H_0,
\end{equation}
where $q_0 = -0.55$ is the deceleration parameter and $j_0 = 1$ is the jerk parameter \cite{2022ApJ...934L...7R}. The corresponding value of $\Omega_m$ is approximately $0.3$, and variations in this parameter have negligible impact on the inferred value of $H_0$.

In the third rung of the distance ladder, systematic uncertainties include: 

\textbf{1. Redshift.} The inference of the Hubble constant relies heavily on redshift measurements. As indicated by Equation (\ref{eq:3}), the measured redshift of an object directly impacts the derived value of $H_0$. Furthermore, redshift is introduced exclusively at the third rung of the distance ladder. Consequently, any systematic uncertainties associated with redshift can lead to biases in the inferred Hubble constant. Studies of bulk flow models in the nearby Universe and corrections for peculiar velocities can affect the inferred value of $H_0$ by up to $0.5$ km s$^{-1}$ Mpc$^{-1}$ \cite{2022ApJ...938..112P,2022PASA...39...46C}.

\begin{figure}[H]
    \includegraphics[width=0.8\textwidth]{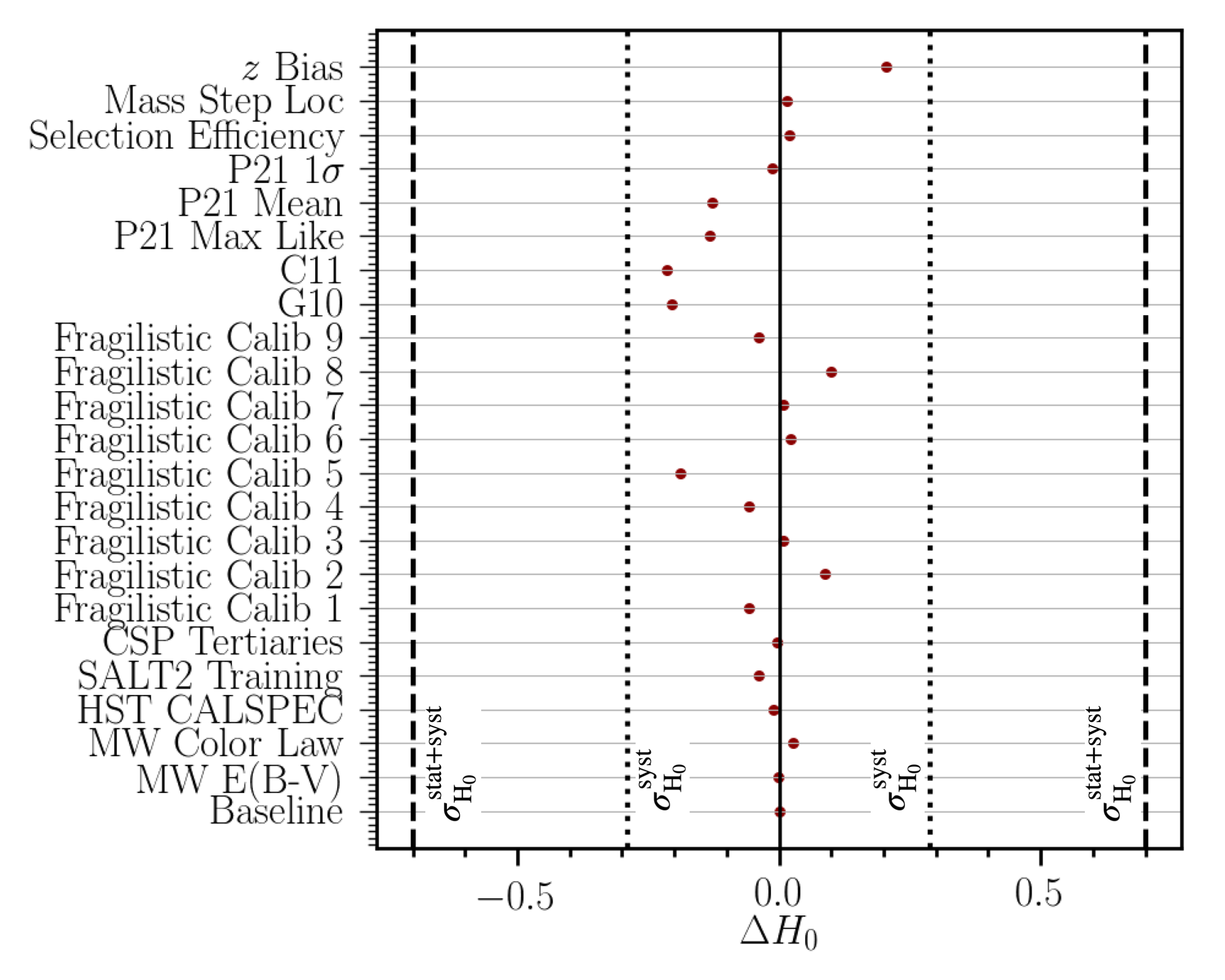}
      \vspace{-6pt}
    \caption{The 
 impact on $H_0$ of the various systematic uncertainties tabulated. The units of these measurements are km s$^{-1}$ Mpc$^{-1}$. More details can be seen in Ref. \cite{Pantheon+}.}
    \label{fig:impact on H0}
\end{figure}

\textbf{2. Dust extinction.} The effects of dust extinction and color-dependent corrections for SNe Ia are absorbed into the nuisance parameter $\beta$. As a cross-check, the parameter was independently fitted for the SNe used in the second and third rungs of the distance ladder, and the resulting values were found to be statistically consistent \cite{Pantheon+}. Figure \ref{fig:impact on H0} summarizes the impact of various systematic uncertainties on $H_0$, adapted from Ref. \cite{Pantheon+}.

\subsection{Tip of the Red Giant Branch}
The TRGB serves as a precise method for distance measurement \cite{1993ApJ...417..553L}. Prior to the onset of the helium flash, red giants reach a peak luminosity that is nearly the same from stat. This allows distances to be derived directly once this luminosity feature is observed. Consequently, the TRGB offers a robust alternative to Cepheids within the distance ladder \cite{2019ApJ...882...34F}. Comparisons between TRGB and Cepheid measurements in SNe host galaxies demonstrate that the two methods yield consistent results \cite{2024ApJ...966...89A,2024ApJ...977..120R}. The TRGB method is typically used to measure $H_0$ by calibrating either the surface brightness fluctuation \cite{2024ApJ...973...83A,2025ApJ...987...87J} or SNe Ia \cite{2019ApJ...882...34F,2022ApJ...932...15A,2023ApJ...954L..31S}. 

Compared to other distance indicators, the TRGB method offers several distinct advantages. Chief among them is its status as a theoretically well-motivated standard candle. This is because the TRGB feature has a clear astrophysical origin: it marks the point of extremely rapid thermonuclear ignition, known as the helium flash, within the degenerate helium cores of low-mass red giant stars \cite{1997MNRAS.289..406S,2002PASP..114..375S}. Second, the measurement of the TRGB is inherently statistical. The apparent magnitude $m_\mathrm{T}$ is determined as the inflection point of the RGB luminosity function \cite{2017ApJ...845..146H,2024ApJ...963L..43A}. Finally, the TRGB is relatively insensitive to metallicity. It is most commonly measured in the $I$ band, where its brightness exhibits minimal variation and is independent of both metallicity and age \cite{1993ApJ...417..553L,2023AJ....166..224M}. However, the TRGB distance measurement method is not without its observational challenges. Small amplitude pulsations near the RGB Tip introduce difficult-to-control systematic uncertainties, typically at the level of a few percent \cite{2024ApJ...974..181K}. Furthermore, RGB populations are susceptible to contamination by other stars, particularly at the bright end. The specifics of such contamination cannot be reliably known from theory alone. 

Several studies have utilized the TRGB as a standard candle to determine $H_0$. Based on the same set of HST $I$ band TRGB observations combined with the Carnegie Supernova Project (CSP) SNIa dataset, \citet{2021ApJ...919...16F} and \citet{2022ApJ...932...15A} adopted discrepant absolute TRGB calibrations and measurement methodologies, yielding $H_0 = 69.8 \pm 0.6\rm{(stat)} \pm 1.6\rm{(sys)}$ km s$^{-1}$ Mpc$^{-1}$ and $H_0 = 71.5 \pm 1.8$ km s$^{-1}$ Mpc$^{-1}$, respectively. By employing JWST $I$ band TRGB measurements alongside the CSP SNe Ia dataset, \mbox{\citet{2025ApJ...985..203F}} derived a Hubble constant of \mbox{$H_0 = 70.39 \pm 1.22 \rm{(stat)} \pm 1.33 \rm{(sys)}$ km s$^{-1}$ Mpc$^{-1}$.} Furthermore, replacing HST entirely with JWST $I$ band TRGB observations of $17$ unique SNe Ia results in $H_0 = 72.1 \pm 2.2$ km s$^{-1}$ Mpc$^{-1}$ \cite{2024ApJ...977..120R}. All these measurements are well above the $\Lambda$CDM prediction.

\subsection{Quasar Gravitational Lensing}
Strong-lensing time delays provide a powerful probe for determining the Hubble constant, independent from other traditional methods. It offers an unparalleled combination of high sensitivity to the Hubble constant and minimal dependence on other cosmological parameters, relying on well understood fundamental physics such as general relativity. The gravitational potential of a lens curves the surrounding spacetime, altering the geometric path length \cite{2002LNP...608.....C}. This effect introduces a time delay $\Delta \tau_{obs}$ in the arrival of the light at the observer. The time delay can be used to infer the corresponding distance as:
\begin{eqnarray}\label{Dt}
	D_{\Delta t} \equiv (1+z_{lens})\frac{D_{d}D_{s}}{D_{ds}},
\end{eqnarray}
where $z_{lens}$ is the lens redshift, $D_{d}$ is the angular diameter distance to the lens, $D_{s}$ is the angular diameter distance to the source, and $D_{ds}$ is the angular diameter distance between the source and the lens, measured at the epoch that the lens is currently observed at \cite{2015MNRAS.450.3155B}. Currently, the precision of the Hubble constant derived from this method is relatively low. The main systematics affecting the precision include: 1. Uncertainties in the lens mass model. This is currently one of the dominant sources of systematics \cite{2013A&A...559A..37S}. Incorporating additional information helps to break the mass-sheet degeneracy. 2. Line-of-sight structures and environmental effects. The distribution of other matter along the light path also produces weak gravitational lensing effects, leading to a biased $H_0$ \cite{2004ApJ...612..660K,2010ApJ...711..201S}. 3. Observational and data processing errors. Although monitoring techniques are mature, the analysis of light curves remains challenging \cite{2015ApJ...800...11L}. Studies of gravitationally lensed quasars have yielded measurements of $H_0$ at different lens redshifts. In a flat $\Lambda$CDM model, a value of $H_0 = 73.3^{+1.7}_{-1.8}$ km s$^{-1}$ Mpc$^{-1}$ is obtained from a joint analysis of six lensed quasars \cite{2020A&A...639A.101M}. There are several estimations derived from the quasar lensing systems shown in Table \ref{table:1}. It is worth noting that $H_0$ values derived from strong gravitational lensing are not exclusively high, with lower values also reported \cite{2023Sci...380.1322K,2024A&A...684L..23G,2025PhRvD.112l3526L}.

\begin{table}[H]
    \newcolumntype{Y}{>{\centering\arraybackslash}X}
    
    \caption{$H_{0}$ 
 estimations derived from the gravitationally lensed quasars. More details can be seen in Ref. \cite{2023Univ....9...94H}.}
    \label{table:1}
    
    \begin{tabularx}{\textwidth}{@{} l Y Y Y l @{}}
        \toprule
        \textbf{Lens Name} & 
        \multicolumn{1}{c}{$\boldsymbol{z_{d}}$} & 
        \multicolumn{1}{c}{$\boldsymbol{z_{S}}$} & 
        \multicolumn{1}{c}{$\boldsymbol{H_{0}}$ \textbf{(km s}\boldmath{$^{-1}$} \textbf{Mpc}\boldmath{$^{-1}$}\textbf{)}} & 
        \textbf{Reference 
} \\
        \midrule
        B1608+656 
    & 0.6304 & 1.394 & $71.0^{+2.9}_{-3.3}$ & \cite{2010AA...524A..94S,2019Sci...365.1134J} \\\midrule
        RXJ1131-1231 & 0.295  & 0.654 & $78.3^{+3.4}_{-3.3}$ & \cite{2014ApJ...788L..35S,2019MNRAS.490.1743C,Shajib2023} \\\midrule
        HE0435-1223  & 0.4546 & 1.693 & $71.7^{+4.8}_{-4.5}$ & \cite{2017MNRAS.465.4895W,2019MNRAS.490.1743C} \\\midrule
        SDSS 1206+4332& 0.745 & 1.789 & $68.9^{+5.4}_{-5.1}$ & \cite{2019MNRAS.484.4726B} \\\midrule
        WFI2033-4723 & 0.6575 & 1.662 & $71.6^{+3.8}_{-4.9}$ & \cite{2020MNRAS.498.1440R} \\\midrule
        PG1115+080   & 0.311  & 1.722 & $81.1^{+8.0}_{-7.1}$ & \cite{2019MNRAS.490.1743C} \\\midrule
        DES J0408-5354& 0.597 & 2.375 & $74.2^{+2.7}_{-3.0}$ & \cite{2017ApJ...838L..15L,2020MNRAS.494.6072S} \\
        \bottomrule
    \end{tabularx}
    
\end{table}

\subsection{Cosmic Chronometers (CCs)}
CCs have recently established themselves as a premier model-independent probe for cosmic expansion \cite{2022LRR....25....6M}. This methodology \cite{2002ApJ...573...37J} yields direct measurements of the Hubble parameter $H(z)$, independent of any assumed cosmological framework. It rests solely on the Cosmological Principle and the validity of General Relativity and standard physics in galactic stellar environments. By analyzing the differential age evolution of the Universe—specifically, the variation in cosmic time $dt$ across a redshift interval $dz$—the method determines $H(z)$ via the relation:
\begin{equation}
    H(z) = -\frac{1}{(1+z)} \frac{dz}{dt}.
\end{equation}
Massive, 
passively evolving galaxies serve as the most effective cosmic chronometers for tracing this age evolution. Characterized by remarkable homogeneity and coeval star formation histories, these systems act as reliable tracers of the oldest stellar populations at any given redshift \cite{2022LRR....25....6M}. Various studies have applied this technique, employing different methodologies to estimate the differential age $dt$ \cite{2015MNRAS.450L..16M,2016JCAP...05..014M,2023JCAP...11..047J,2025A&A...703A.232L,2026A&A...707A.111T,2026arXiv260304872V}. Currently, the available sample includes over 30 independent $H(z)$ measurements across the range $0<z<2.1$, with uncertainties varying from approximately $5\%$ at low redshifts to $10$--$15\%$ 
 at high redshifts. Several systematic uncertainties exist within this method, such as residual contamination from young, star-forming outliers, the age-metallicity degeneracy, and the dependence on assumed Stellar Population Synthesis models to derive stellar ages. Additionally, there are other effects that may potentially bias the age--redshift relation. \citet{2020ApJ...898...82M} proposed using a covariance matrix to account for the impact of these systematics, though it seems likely that the uncertainties are overestimated \cite{2025JCAP...10..041K}. For a comprehensive reviews of the method and its applications, see Refs. \cite{2022LRR....25....6M,2023arXiv230709501M,2024arXiv241201994M}.

Because CCs deliver direct measurements of the Hubble parameter without assuming a specific cosmological model, they have emerged as a vital instrument in Hubble tension research. To maximize their utility, researchers have applied various cosmological model-independent reconstruction techniques to CC data. These include Gaussian Processes \mbox{(GPs) \cite{2018ApJ...856....3Y,2021arXiv210108565C,2024EPJC...84....3Y,2024JHEAp..44..323F},} non-parametric smoothing methods \cite{2017JCAP...03..005R,2016PhRvD..93d3014L}, Weighted Polynomial Regression \cite{2018JCAP...04..051G}, Padé approximations \cite{2022ApJ...939...37L}, and Artificial Neural Network (ANN) architectures \cite{2024EPJC...84....3Y}. All such methods facilitate a cosmology-free reconstruction of $H(z)$, allowing for extrapolation to redshift $z=0$ to derive the present-day Hubble constant. Conversely, one may adopt a specific cosmological model to constrain parameters via parametric fitting \cite{2012JCAP...07..053M,2016JCAP...12..039M,2022LRR....25....6M,2024MNRAS.527.4874C,2025ApJ...978L..33G}. Remarkably, provided that statistical and systematic uncertainties are properly handled, both model-independent and model-dependent analyses yield consistent outcomes, specifically $H_0 = 70.7 \pm 6.7$ km s$^{-1}$ Mpc$^{-1}$ \cite{2023MNRAS.523.3406F}.

\subsection{Fast Radio Bursts (FRBs)}
FRBs are millisecond-duration pulses occurring at cosmological distances \cite{Lorimer2007,Xiao2021,Zhang2022}. Because the propagation speed of electromagnetic waves in a plasma is frequency-dependent, a signal emitted instantaneously at the source reaches Earth with its lower-frequency components arriving later than the higher-frequency ones—a phenomenon known as dispersion. The observed dispersion measure (DM) is an important observational parameter for FRBs. It is typically expressed in the following form:
\begin{equation}
    \mathrm{DM}_{\mathrm{obs}} = \mathrm{DM}_{\mathrm{MW}} + \mathrm{DM}_{\mathrm{halo}} + \mathrm{DM}_{\mathrm{IGM}} + \frac{\mathrm{DM}_{\mathrm{host}}}{1+z}.\label{eq:DM}
\end{equation}
Here, $\mathrm{DM}_{\mathrm{obs}}$ is the total observed DM, and $\mathrm{DM}_{\mathrm{host}}$ is the DM of the host galaxy in the FRB's source frame, where the factor of $(1+z)$ accounts for time dilation. $\mathrm{DM}_{\mathrm{MW}}$, the DM contributed by the interstellar medium of the MW, is well described by Galactic electron distribution models such as YMW16 and NE2001 \citep{2002astro.ph..7156C,2017ApJ...835...29Y}. The component $\mathrm{DM}_{\mathrm{halo}}$ is currently poorly constrained. It is commonly assumed to follow a Gaussian distribution with $\langle \mathrm{DM}_{\mathrm{halo}} \rangle$ =
$65~\mathrm{pc~cm^{-3}}$ and $\sigma = 15~\mathrm{pc~cm^{-3}}$ \citep{2019MNRAS.485..648P,2020Natur.581..391M,2022MNRAS.515L...1W}. Within the framework of the $\Lambda$CDM cosmological model, the mean value of $\mathrm{DM}_{\mathrm{IGM}}$ can be expressed as 
\begin{equation}
    \langle \mathrm{DM}_{\mathrm{IGM}} \rangle = \frac{3cH_0\Omega_b f_{\mathrm{IGM}}}{8\pi G m_p} \times f_e(z),
\end{equation}
where $c$ is the speed of light, $H_0$ is the Hubble constant, $\Omega_b$ is the density of baryon matter, $f_{\mathrm{IGM}}$ is the fraction of baryons in IGM, $G$ is the Newtonian gravitational constant, $m_p$ is the proton mass, and $f_e(z)$ is defined as 
\begin{equation}
    f_e(z) = \int_0^z \frac{\left[ \frac{3}{4} y_1 \chi_{e,\mathrm{H}}(z) + \frac{1}{8} y_2 \chi_{e,\mathrm{He}}(z) \right] (1+z)\,dz}{\left[ \Omega_m (1+z)^3 + \Omega_\Lambda \right]^{1/2}}.
\end{equation}
The parameters $y_1$ and $y_2$ are the hydrogen and helium fractions normalized to $0.76$ and $0.24$, respectively, which can be neglected as $y_1 \simeq y_2 \simeq 1$. In the late universe ($z<3$), hydrogen and helium can be considered fully ionized \citep{2009RvMP...81.1405M,2011MNRAS.410.1096B}. Therefore, their ionization fractions $\chi_{e,\mathrm{H}}(z)$ and $\chi_{e,\mathrm{He}}(z)$ can be taken as $\chi_{e,\mathrm{H}}(z) = \chi_{e,\mathrm{He}}(z) = 1$. Finally, the expression for $\mathrm{DM}_{\mathrm{IGM}}$ can be simplified to
\begin{equation}
    \langle \mathrm{DM}_{\mathrm{IGM}} \rangle = \frac{21 c \Omega_b H_0^2}{64 \pi H_0 G m_p} \times \int_0^z \frac{f_{\mathrm{IGM}} (1+z)\,dz}{\left[ \Omega_m (1+z)^3 + 1 - \Omega_m \right]^{1/2}}.\label{eq:DM_IGM}
\end{equation}
The value of $f_{\mathrm{IGM}} = 0.84$ is typically adopted \cite{2012ApJ...759...23S,Yang2022}. 
Then, the dispersion measure--redshift relation allows FRBs to be used as cosmological probes \cite{2025A&A...698A.215G}. 
However, the degeneracy between $\mathrm{DM}_{\mathrm{IGM}}$ and $\mathrm{DM}_{\mathrm{host}}$ is the main obstacle for the cosmological application of FRBs. Early studies assumed fixed values for the dispersion measures contributed by $\mathrm{DM}_{\mathrm{IGM}}$ and $\mathrm{DM}_{\mathrm{host}}$, despite the practical difficulty in distinguishing between the two. A reasonable approach is to consider the probability distributions of $\mathrm{DM}_{\mathrm{IGM}}$ \cite{2014ApJ...780L..33M,Zhangzj2021} and $\mathrm{DM}_{\mathrm{host}}$ \cite{Zhanggq2020}. The best-fit distribution parameters of $\mathrm{DM}_{\mathrm{IGM}}$ and $\mathrm{DM}_{\mathrm{host}}$ are derived from the IllustrisTNG simulation \cite{Zhanggq2020,Zhangzj2021}.

\citet{2022MNRAS.511..662H} presented the first estimation of the Hubble constant, $H_{0} = 62.3 \pm 9.1\,\mathrm{km\,s^{-1}\,Mpc^{-1}}$, based on a sample of $9$ FRBs. By incorporating priors for \mbox{$\mathrm{DM}_{\mathrm{IGM}}$ \cite{Zhangzj2021}} and $\mathrm{DM}_{\mathrm{host}}$ \cite{Zhanggq2020} derived from the IllustrisTNG simulation, \citet{2022MNRAS.515L...1W} obtained a more constrained result of $H_{0} = 68.81^{+4.99}_{-4.33}\,\mathrm{km\,s^{-1}\,Mpc^{-1}}$ using $18$ localized FRBs. Subsequently, \citet{2022MNRAS.516.4862J} reported $H_{0} = 73^{+12}_{-8}\,\mathrm{km\,s^{-1}\,Mpc^{-1}}$. Their analysis employed an updated sample comprising 16 Australian Square Kilometre Array Pathfinder (ASKAP) FRBs detected by the Commensal Real-time ASKAP Fast Transients (CRAFT) Survey and localized to their host galaxies, and 60 unlocalized FRBs from Parkes and ASKAP. Furthermore, \citet{2022MNRAS.511..662H} conducted a forecast using a realistic mock sample, demonstrating that high-precision measurements of the expansion rate are achievable without relying on external cosmological probes. Adopting a cosmological-model-independent method, \citet{2022arXiv221005202L} derived $H_{0} = 70.60 \pm 2.11\,\mathrm{km\,s^{-1}\,Mpc^{-1}}$. More recently, \citet{2022arXiv221213433Z} performed the first statistical $H_{0}$ measurements using unlocalized FRBs, yielding $H_{0} = 71.7^{+8.8}_{-7.4}\,\mathrm{km\,s^{-1}\,Mpc^{-1}}$ and $H_{0} = 71.5^{+10.0}_{-8.1}\,\mathrm{km\,s^{-1}\,Mpc^{-1}}$ for simulation-based and observation-based scenarios, respectively. Finally, \citet{2025A&A...698A.215G} utilized a sample of 108 localized FRBs to derive $H_0 = 69.4^{+2.14}_{-1.97}\,\mathrm{km\,s^{-1}\,Mpc^{-1}}$. Additionally, by applying Monte Carlo methods to 527 non-localized FRBs, they constrained the Hubble constant to $H_0 = 68.81 \pm 0.68\,\mathrm{km\,s^{-1}\,Mpc^{-1}}$. Given that these unlocalized FRBs are typically at $z \gtrsim 0.2$, this low value for $H_0$ suggests that the Hubble tension is not present in the high-\mbox{redshift Universe}.

\section{Solutions to the Hubble Tension} \label{Sec:Explanations for the Hubble tension}
\subsection{Redshift Dependence of Parameters}
\begin{figure}[H]
    \includegraphics[width=0.8\textwidth]{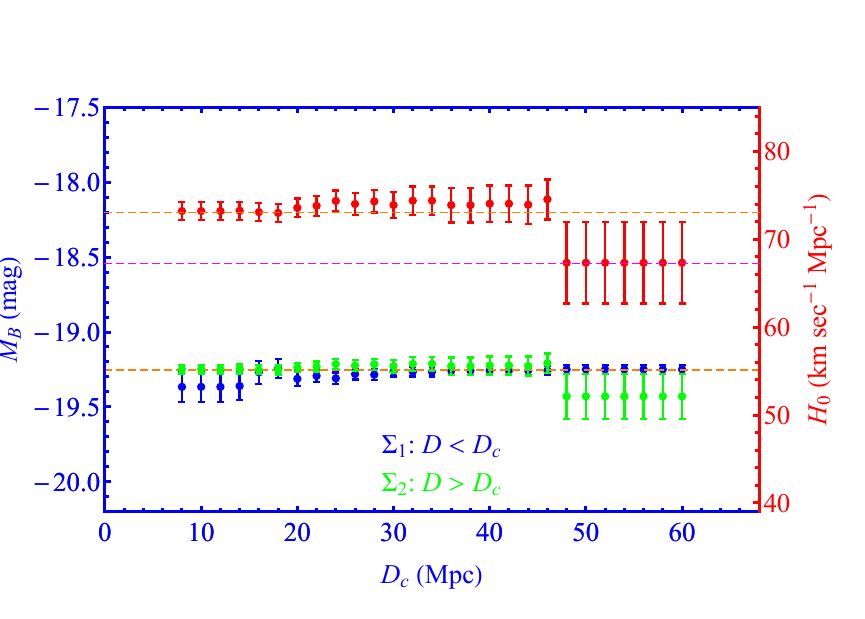}
    \vspace{-5pt}
    \caption{The 
best fit values with uncertainties for the SNe Ia absolute magnitude $M_B$ and for $H_0$, derived from SNe within and beyond the transition distance $D_c$. More details can be seen in \mbox{Ref. \cite{2022Univ....8..502P}.}}
    \label{fig:transition Mb}
\end{figure}

To address the Hubble tension, researchers have explored numerous avenues, one of which involves scrutinizing the evolution and behavior of key cosmological parameters. For instance, a recent reanalysis of the SH0ES data tested the homogeneity of the Cepheid-SNe Ia distance ladder by allowing individual modeling parameters to undergo a discrete transition at a critical distance of approximately $50$ Mpc \cite{2022Univ....8..502P}. The study found that permitting the SN Ia absolute magnitude $M_B$ to shift beyond this scale yields a best-fit Hubble constant of $H_0 = 67.32 \pm 4.64$ km s$^{-1}$ Mpc$^{-1}$, fully consistent with the Planck value, and strongly favors this transitional model over the standard SH0ES baseline when combined with inverse distance ladder constraints. When allowing the absolute magnitude $M_B$ of SNe Ia to vary, the best-fit value of $M_B$ and the resulting inferred Hubble constant $H_0$ as a function of distance $D$ are shown in Figure \ref{fig:transition Mb}. The inferred transition aligns with the shift needed to resolve the Hubble tension, assuming a fundamental change in \mbox{gravity---specifically,} a sudden enhancement in the effective gravitational coupling $G_{eff}$ at redshift $z_{t}$ $\lesssim 0.01$ \cite{2021PhRvD.104b1303M}. Such a gravitational transition would lead to an abrupt brightening of SNe Ia by approximately $\Delta$$M_{B}$ $\backsimeq$ 0.2 \cite{2021PhRvD.104b1303M,2021PhRvD.103h3517A}. However, this scenario encounters very severe problems when confronted with terrestrial and Solar System constraints from the corresponding look back time \cite{2025MNRAS.539.1553B}.

In addition, there have been studies focusing on the standardization relations used for SNe Ia. For example, recent work \cite{2025arXiv251220834P} re-examined the well-known “mass step”—the correlation between standardized SN Ia brightness and host galaxy stellar mass—using improved stellar mass estimates derived from optical to mid-infrared photometry and nonparametric star formation histories. The study suggests that the trend is likely driven primarily by environmental metallicity: Type Ia supernovae in low-metallicity host galaxies appear systematically fainter, indicating that metallicity—rather than stellar mass itself—plays a central role in determining their standardized luminosities. Separately, another investigation has uncovered a subtle but persistent correlation between locally measured values of $H_0$ and the large-scale ambient density environment of SN Ia host galaxies \cite{2022arXiv220914732Y}. By analyzing SNe Ia beyond the homogeneity scale to minimize sample variance, the authors identified a residual linear trend: $H_0$ inferred from different SN subsamples correlates with the density contrast of their host environments, with the correlation strengthening at larger scales—contrary to, though still marginally consistent with, $\Lambda$CDM predictions. This unexpected scale dependence may hint at unaccounted-for peculiar velocity corrections or, more speculatively, new physics related to the growth of structure and the resulting peculiar velocities on large scales \cite{2023MNRAS.524.1885W,2025arXiv251203168W}. 

\subsection{Statistical Methods}
\begin{figure}[H]
    \includegraphics[width=0.8\textwidth]{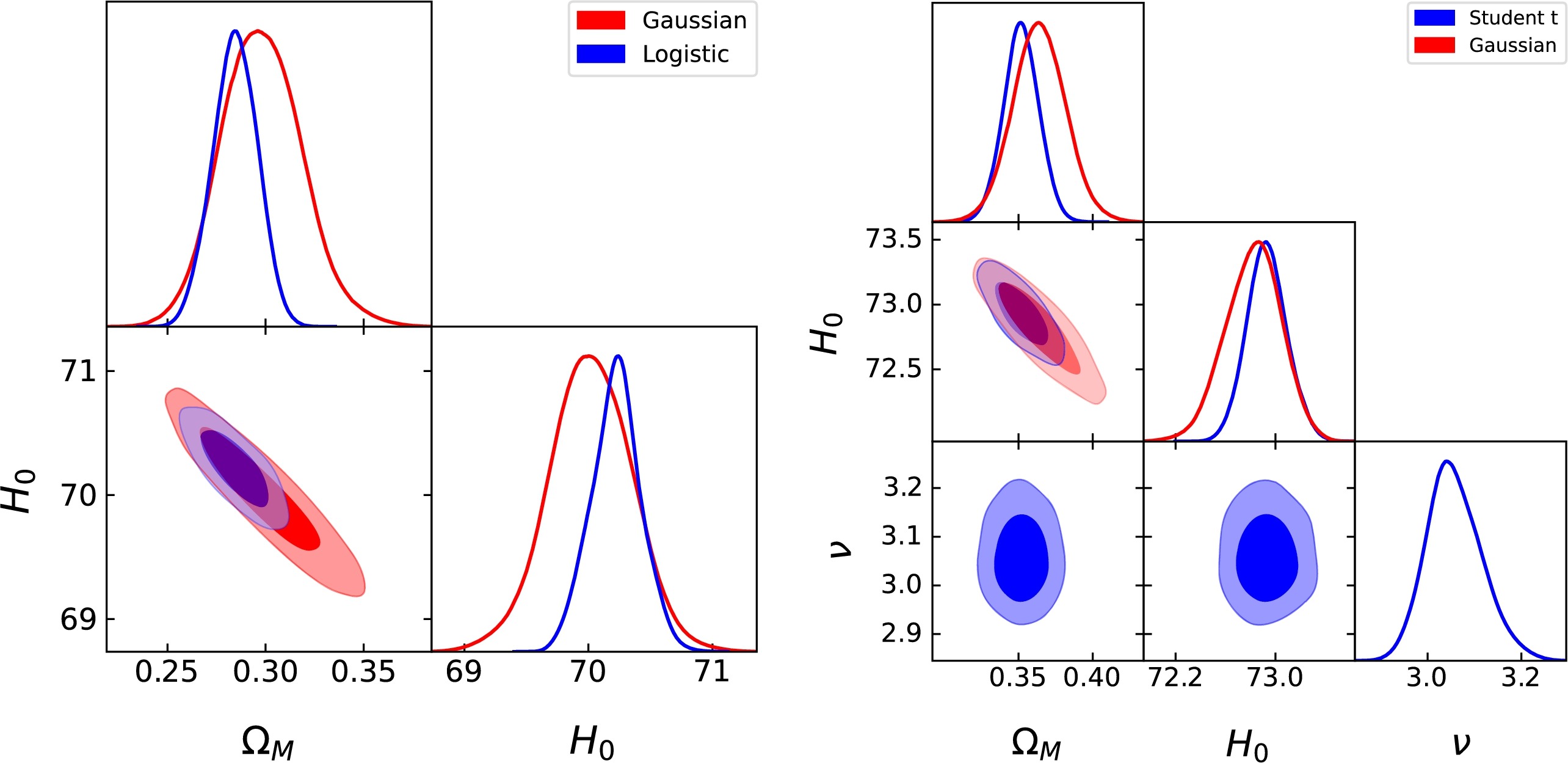}
    \caption{Fit 
of the flat $\Lambda$CDM model with $\Omega_m$ and $H_0$ as free parameters. More details can be seen in Ref. \cite{2024JHEAp..41...30D}.}
    \label{fig:7}
\end{figure}

When observational data alone prove insufficient to resolve the current Hubble tension, some researchers have turned to statistical methodologies as a potential avenue for reconciliation. Given that the determination of the Hubble constant relies on a large volume of observational data and employs statistical tools---such as Bayesian inference---in the fitting process, a statistical interpretation of the Hubble tension is well justified. When fitting SNe Ia data, the likelihood function is typically assumed to be Gaussian as:
\begin{equation}
    \log \mathcal{L} = -\frac{1}{2} \left( \Delta^T C^{-1} \Delta + \log |2\pi C| \right),
\end{equation}
where $C$ is the covariance matrix and $\Delta$ is the residual. The likelihood function commonly assumes that the residuals follow a unit Gaussian; however, real observational samples deviate from this assumption and do not strictly conform to a Gaussian form \cite{2025MNRAS.536..234L}. By exploring alternative likelihood functions, although the inferred value of $H_0$ is not significantly reduced, the precision of cosmological parameter constraints is notably \mbox{improved \cite{2025MNRAS.536..234L,2024JHEAp..41...30D},} as shown in Figure \ref{fig:7}. Moreover, studies on the choice of prior assumptions have also indicated that they may introduce bias \cite{2025arXiv251103394D}. It highlights the crucial need to model both the distance prior and selection accurately for robust distance ladders and \mbox{derived parameters.}

\subsection{Theoretical Models}
The analyses of systematics have so far proven insufficient to fully account for the current Hubble tension, suggesting that it may arise from new physics beyond the standard cosmological model \cite{2020PhRvD.102b3518V}. To date, numerous theories have been proposed to attempt to resolve or alleviate the Hubble tension, including \cite{2017PhRvD..96d3503D,2018PhRvD..97d3528D,2019ApJ...883L...3L,2019JCAP...12..035K,2019PDU....2600385D,2019PhRvD.100d3537D,2020PhRvD.101j3517B,2020PhRvD.102b3518V,2021Entrp..23..404D,2021PhRvD.103h3517A,2021ApJ...923..212B,2023PhRvD.108l4050H,2024PhRvD.109b3527A}, and reviews such \mbox{as \cite{2021CQGra..38o3001D,2021A&ARv..29....9S,2022NewAR..9501659P,2025PDU....4901965D}.} Recent observations showing no discrepancies in CMB anisotropies under the $\Lambda$CDM model serve as a strong argument against early-time resolutions \cite{2025JCAP...11..063C,2025arXiv250620707C}. Furthermore, models that substantially increase the $H_0$ value required to fit the CMB usually entail an elevated baryon density, thereby exacerbating tensions with Big Bang nucleosynthesis \cite{2026arXiv260405095G,2026arXiv260416600L}. Analyses of quasar lensing suggest that the Hubble constant inferred from strongly lensed quasar time delays (H0LiCOW) exhibits a mild decline with increasing lens redshift \cite{2020MNRAS.498.1420W}. This trend is statistically modest, with a significance of approximately $1.9 \sigma$. However, with the inclusion of a new $H_0$ measurement \cite{2020MNRAS.494.6072S}, the significance has decreased to $1.7 \sigma$ \cite{2020A&A...639A.101M}. Although a decreasing trend in $H_0$ has been observed via strongly lensed quasars, it is important to note that time-delay cosmography is a relatively novel probe compared to the well-established observations of SNe Ia. Consequently, the current sample size is limited, and potential systematic uncertainties remain a concern. Despite the diluted significance of this trend with additional data, it nonetheless offers a fresh perspective on the Hubble tension.

\begin{figure}[H]
    \includegraphics[width=0.8\textwidth]{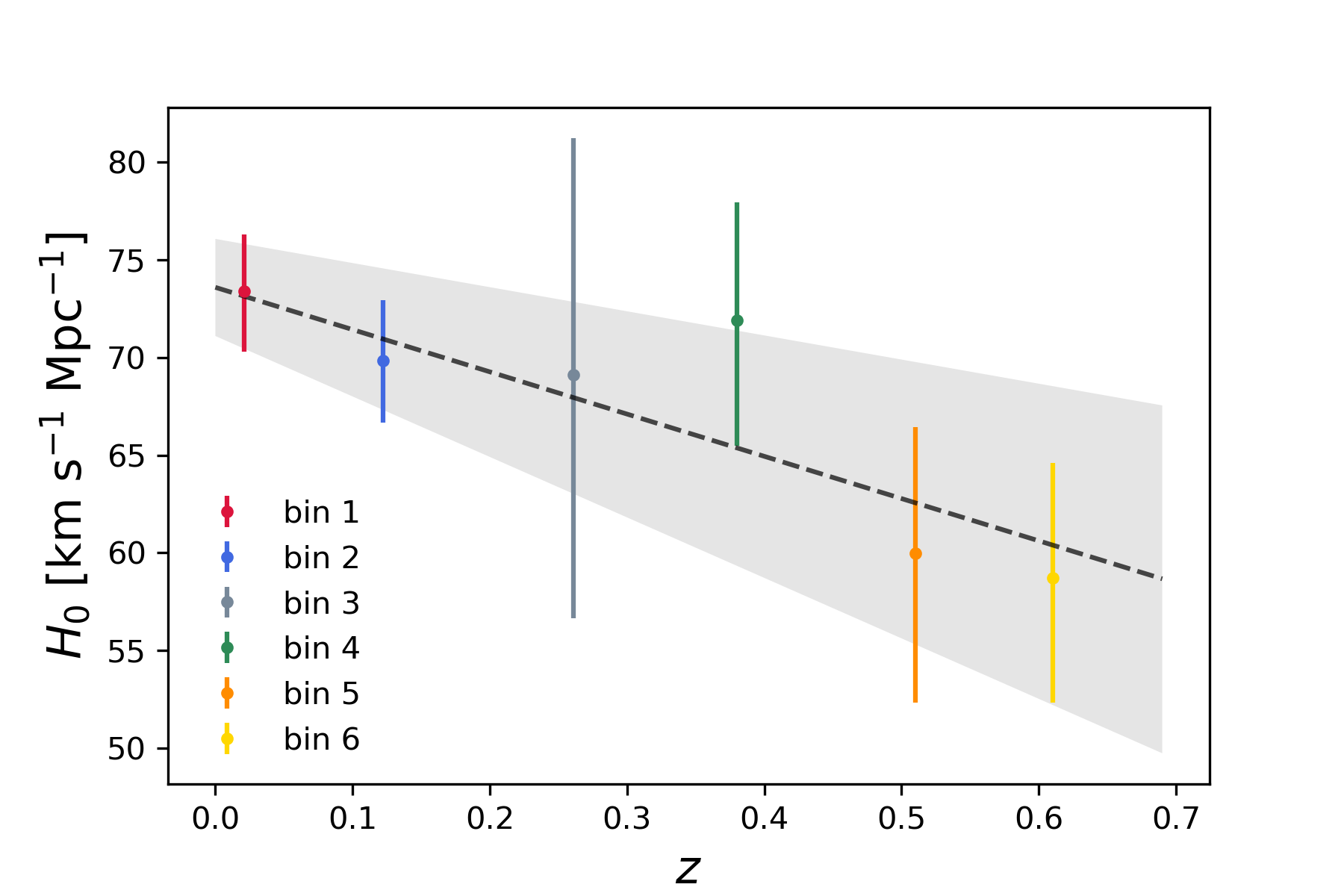}
    \caption{The 
 best-fit results for the binned data. The decline trend is $2.1 \sigma$ significance. More details can be seen in Ref. \cite{2020PhRvD.102j3525K}.}
    \label{fig:8}
\end{figure}

\citet{2020PhRvD.102j3525K} grouped the observational data, including megamasers, CC, SNe Ia, and BAO, into six different redshift bins. The effective redshift for each bin was computed as 
\begin{equation}
    \bar{z} = \frac{\sum_n^{N_i} z_n (\sigma_n)^{-2}}{\sum_n^{N_i} (\sigma_n)^{-2}},
\end{equation}
where $\sigma_n$ denotes the error in the observable at redshift $z_n$. A similar declining trend of $H_0$ is found with a low statistical significance ($2.1 \sigma$), as illustrated in Figure \ref{fig:8}. From this, it is evident that these two independent studies mutually corroborate each other's findings, lending greater credibility to the hypothesis that $H_0$ exhibits a redshift-dependent decline and thereby attracting increased attention to this possibility. 

Building upon the $\Lambda$CDM and $w$CDM cosmological frameworks, \citet{2021ApJ...912..150D} performed fits under the assumption that the Hubble constant evolves with redshift according to a functional form $g(z)$ as:
\begin{equation}
    g(z) = H_0(z) = \frac{\tilde{H}_0}{(1+z)^a},
\end{equation}
where $\tilde{H}_0$ and $\alpha$ are free parameters, with $\alpha$ indicating the evolutionary trend. They explored various binning schemes and ultimately concluded that the declining trend in $H_0$ is robust, persisting independently of both cosmological model and the choice of binning method. Within the $\Lambda$CDM framework, they divided the sample into four redshift bins and obtained a best-fit result of $H_0 = 73.493 \pm 0.144$ km s$^{-1}$ Mpc$^{-1}$ and $\alpha = 0.008 \pm 0.006$. By assuming the $g(z)$ function, they obtained a revised value of \mbox{$H_0(z = 1100) = 69.271 \pm 2.815$ km s$^{-1}$ Mpc$^{-1}$} that alleviates the Hubble tension by approximately $66\%$. A similar declining trend has subsequently been identified in a series of follow-up studies \cite{2022A&A...668A..34H,2022PhRvD.106d1301O,2022Galax..10...24D,2024PDU....4401464O,Dai:2026pvx}.

\begin{figure}[H]
\includegraphics[width=0.8\textwidth]{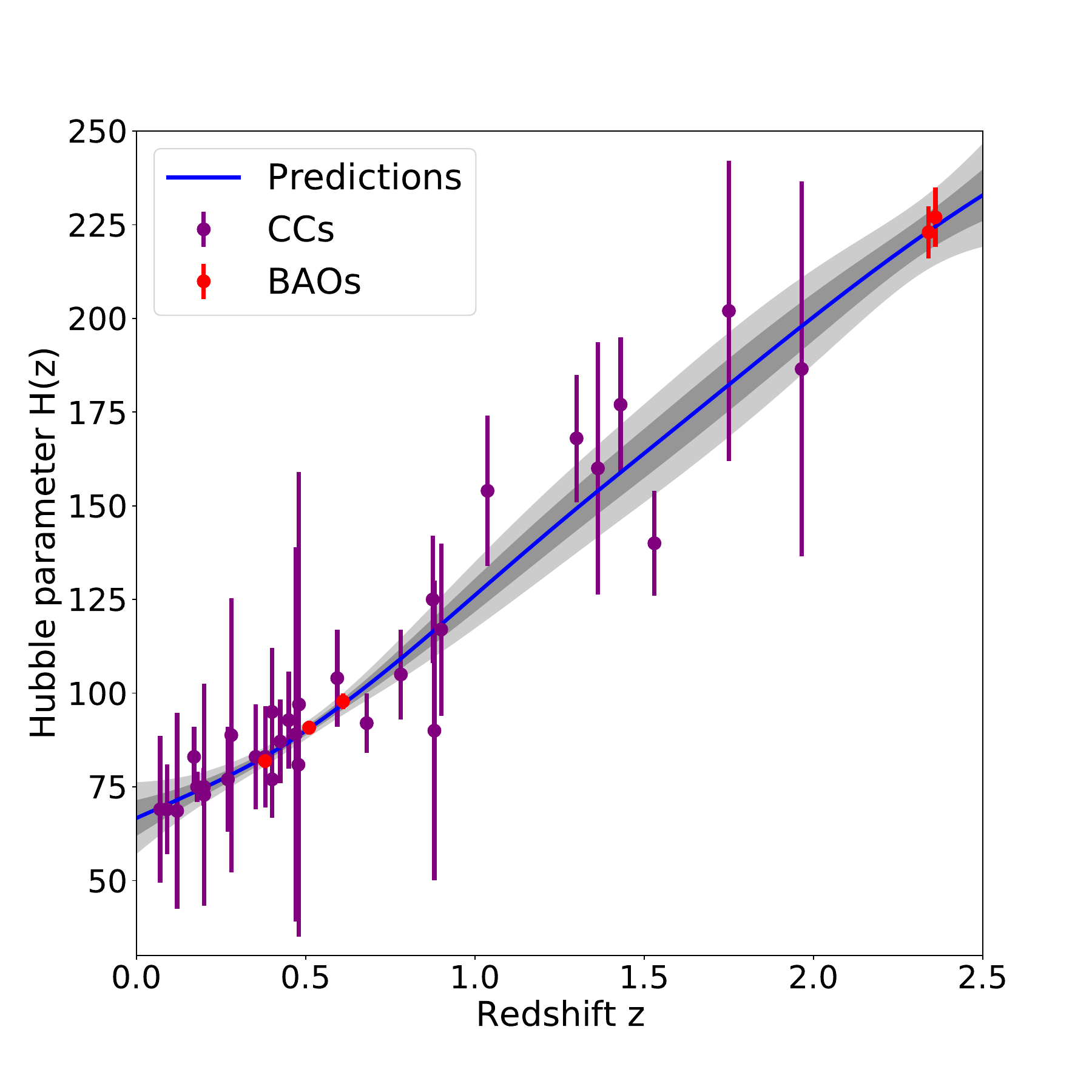}
\vspace{-3pt}
\caption{Smoothed 
 $H(z)$ function (blue solid line) with 2$\sigma$ errors (gray regions) obtained from the 36~H(z) data (31 CC + 5 BAO) employing GP method. More details can be seen in Ref. \cite{2021MNRAS.507..730H}. 
\label{fig:9}}
\end{figure}

\begin{figure}[H]%
\hspace{-2mm}\includegraphics[width=6.8 cm]{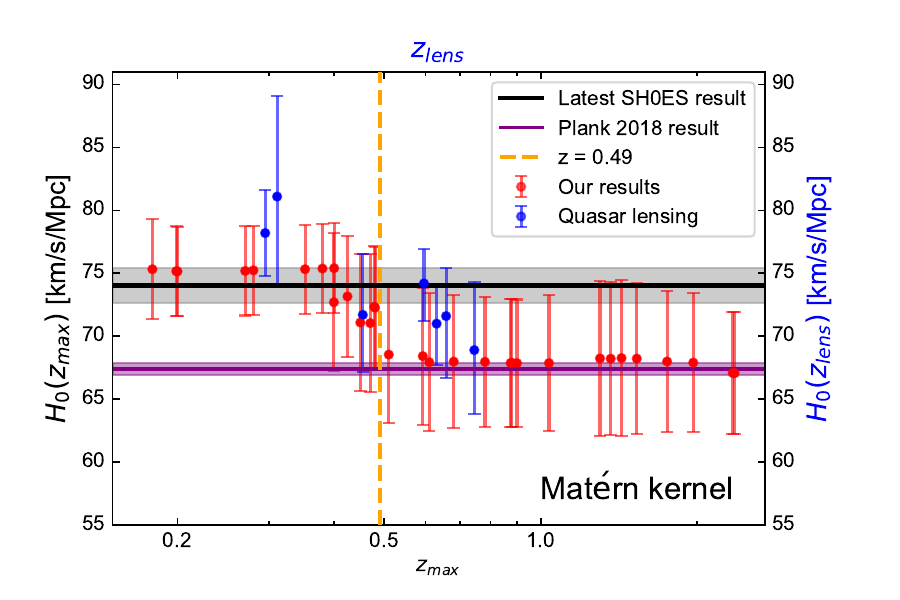}
\includegraphics[width=6.8 cm]{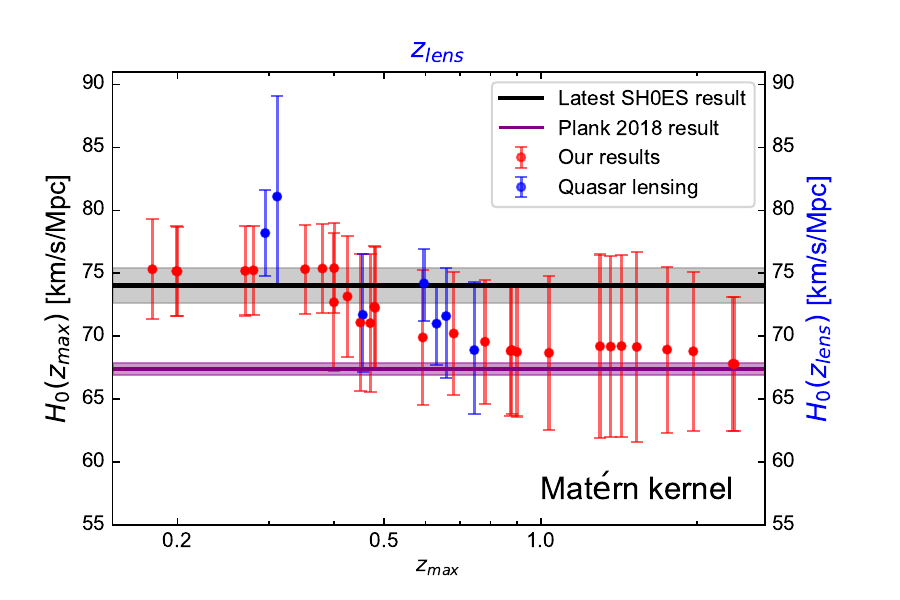}
\vspace{-5pt}
\caption{Predictions of $H_{0}(z_{\rm max})$ derived from a sample of 36 $H(z)$ measurements (31 CC + 5 BAO). Here, $H_{0}(z_{\rm max})$ denotes the Hubble constant inferred from a dataset truncated at a maximum redshift $z_{\rm max}$. The red points represent the predicted values of $H_{0}(z_{\rm max})$ obtained from our analysis. The gray and purple shaded regions indicate the constraints reported by the SH0ES and \emph{Planck} collaborations, respectively. The blue dotted vertical line marks the transition redshift at $z = 0.49$. Additionally, the blue points display the $H_0$ measurements derived from quasar lensing observations, plotted in the ($z_{\rm lens}$, $H_{0}(z_{\rm lens})$) plane. More details can be seen in Ref. \cite{2022MNRAS.517..576H}.
\label{fig:10}}
\end{figure}

Using Gaussian process reconstruction, \citet{2021PhRvD.103j3509K} proposed a diagnostic $\mathcal{H}0(z)$ to probe deviations from the flat $\Lambda$CDM model, where the redshift-dependent diagnostic $\mathcal{H}0(z)$ is expressed as:
\begin{equation}\label{eq:H0D}
\mathcal{H}0(z) = \frac{H(z)}{\sqrt{1-\Omega_{m0} + \Omega_{m0}(1 + z)^{3}}}.
\end{equation}
As a model-independent statistical technique, Gaussian process has been widely employed in cosmological studies \cite{2018JCAP...04..051G,2018ApJ...856....3Y,2019ApJ...886L..23L,2020ApJ...895L..29L,2022PhyS...97h5011S}. Using $H(z)$ data and Gaussian process, \mbox{\citet{2021MNRAS.507..730H}} performed a cosmology-independent reconstruction of the Hubble parameter, as shown in Figure \ref{fig:9}. From Figure \ref{fig:9}, it is evident that combining $H(z)$ data with the Gaussian process method yields a smooth, continuous function $f(z)$ that effectively interpolates the discrete measurements. This reconstructed function enables the estimation of $H(z)$ at any redshift within the coverage of the data---including the present-day value $H_0$. The reliable interpolation range is primarily determined by the highest redshift probed by the $H(z)$ dataset. In this analysis, the reconstructed $H(z)$ is derived from $f(z)$, while the parameter $\Omega_m$ is fixed to the Planck value \cite{2020A&A...641A...6P}. Unlike previous work, \mbox{\citet{2022MNRAS.517..576H}} introduced a new binning approach and performed a Gaussian process reconstruction using 
$H(z)$ data to trace its evolutionary trend. The results reveal a transition $H_0$ during the late Universe. As redshift increases, $H_0$ decreases from a high value at low redshift to a low one at high redshift---a trend that could help explain the Hubble tension. This redshift-dependent evolution of $H_0$ is illustrated in Figure \ref{fig:10}, where the transition occurs at approximately $z \sim 0.49$. Without introducing a new cosmological model, their result offers a resolution to the Hubble tension with a mitigation level of approximately $70\%$, and remains consistent with the H0LiCOW measurement within $1\sigma$ confidence level.

\begin{figure}[H]
\includegraphics[width=0.8\textwidth]{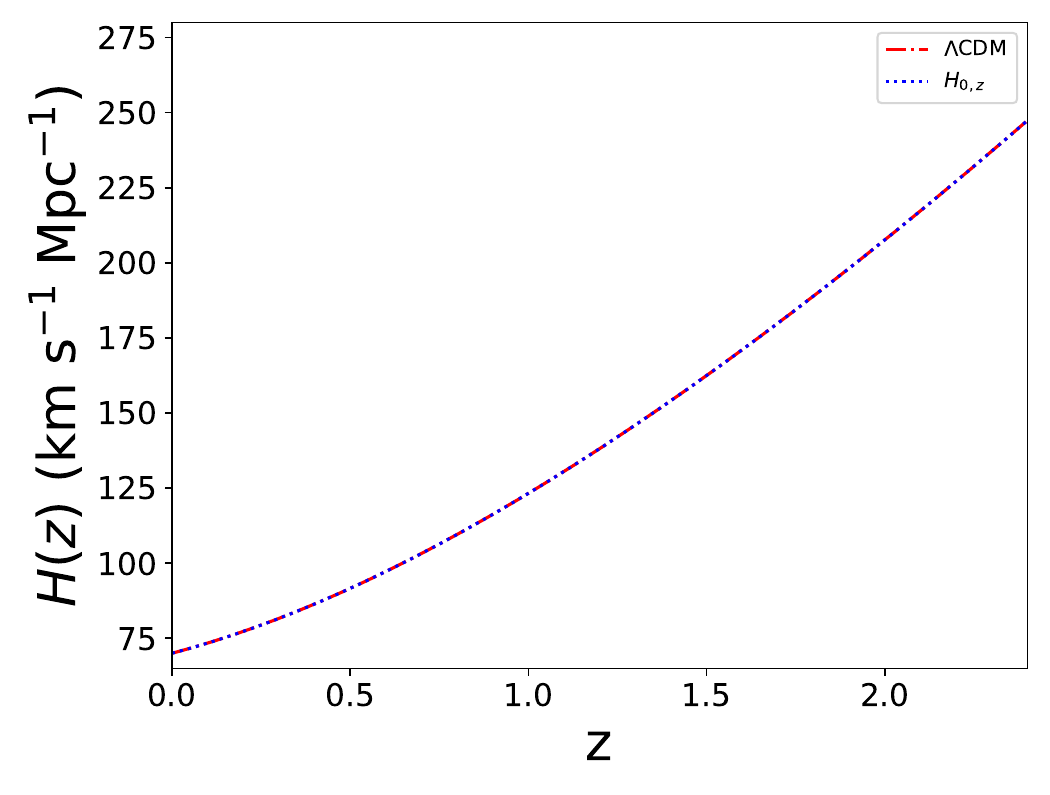}
\vspace{-9pt}
\caption{Comparison between the Hubble parameter $H(z)$ in the standard $\Lambda$CDM model and that derived from Equation (\ref{eq:Hz step}). $H_0=H_{0,z_i} = 70$ km s$^{-1}$ Mpc$^{-1}$, $\Omega_{k0}=0,$ and $\Omega_{m0} = 0.3$ are assumed. More details can be seen in Ref. \cite{2023A&A...674A..45J}. 
\label{fig:11}}
\end{figure}

Following investigations into whether $H_0$ evolves, some studies have taken the next step by examining the specific form or functional behavior of its evolution \cite{2023A&A...674A..45J,2025ApJ...979L..34J}. Inferring the evolution of $H_0$ directly from observational data represents a promising and model-agnostic approach. However, if one merely partitions the dataset and performs separate fits in each bin, the resulting $H_0$ estimates tend to be degenerate with one another. Assuming a specific cosmological model or a prescribed evolution of $H_0$ inevitably introduces additional prior dependence into the results, potentially biasing the inference. A central challenge in this research is obtaining unbiased and mutually uncorrelated estimates of $H_0$ at different redshifts. To address the aforementioned issues, \citet{2023A&A...674A..45J} pioneered the use of a non-parametric approach to infer the values of $H_0$ at different redshifts. Inspired by approaches used to study the equation of state of dark energy \cite{2005PhRvD..71b3506H}, they proposed a similar methodology to reconstruct the evolution of $H_0$. This method is based on the assumption that a varying $H_0(z)$ can be approximated by a stepwise function, which takes a constant value within each redshift interval. This means that the function $H_0(z)$ can be expressed as:
\begin{equation}\label{eq:H0function}
H_0(z)=\begin{cases} H_{0,z_1} &\text{ if } 0\le z < z_1, \\ 
H_{0,z_2} &\text{ if } z_1 \le z < z_2,\\
\cdots  &\cdots,\\
H_{0,z_i} &\text{ if } z_{i-1} \le z < z_i,\\
\cdots &\cdots, \\
H_{0,z_N} &\text{ if } z_{N-1} \le z < z_N.
\end{cases}
\end{equation}
The parameter $i$ means the $i$th redshift bin, $N$ is the number of total redshift bins, and $H_{0,z_i}$ represents the value of $H_0(z)$ between $z_{i-1}$ to $z_i$. By combining this parameterization with a cosmological model, one can obtain an extended evolutionary form of $H_0(z)$ within that framework. The simplest case arises in the standard cosmological model, where $H(z)$ can be expressed as:
\newpage
\vspace*{-21pt}
\begin{equation}\label{eq:Hz step}
\begin{split}
H(z_i)=& H_{0,z_1}\int_{0}^{z_1} \frac{3\Omega_{m0}(1+z)^2}{2\sqrt{\Omega_{m0}(1+z)^{3}+\Omega_{\Lambda0}}} \\
& +H_{0,z_2}\int_{z_1}^{z_2} \frac{3\Omega_{m0}(1+z)^2}{2\sqrt{\Omega_{m0}(1+z)^{3}+\Omega_{\Lambda0}}} \\
& +\cdots \\
& +H_{0,z_i}\int_{z_{i-1}}^{z_i} \frac{3\Omega_{m0}(1+z)^2}{2\sqrt{\Omega_{m0}(1+z)^{3}+\Omega_{\Lambda0}}}+H_{0,z_i}.
\end{split}
\end{equation}
If $H_0(z)$ does not exhibit any evolutionary trend, the result will revert to $H_0$, as shown in Figure \ref{fig:11}. It is evident that when $H_0$ is constant, the $H(z)$ derived from Equation (\ref{eq:Hz step}) exactly matches the prediction of the $\Lambda$CDM model. 

\begin{figure}[H]
\hspace{-4mm}\includegraphics[width=\textwidth,angle=0]{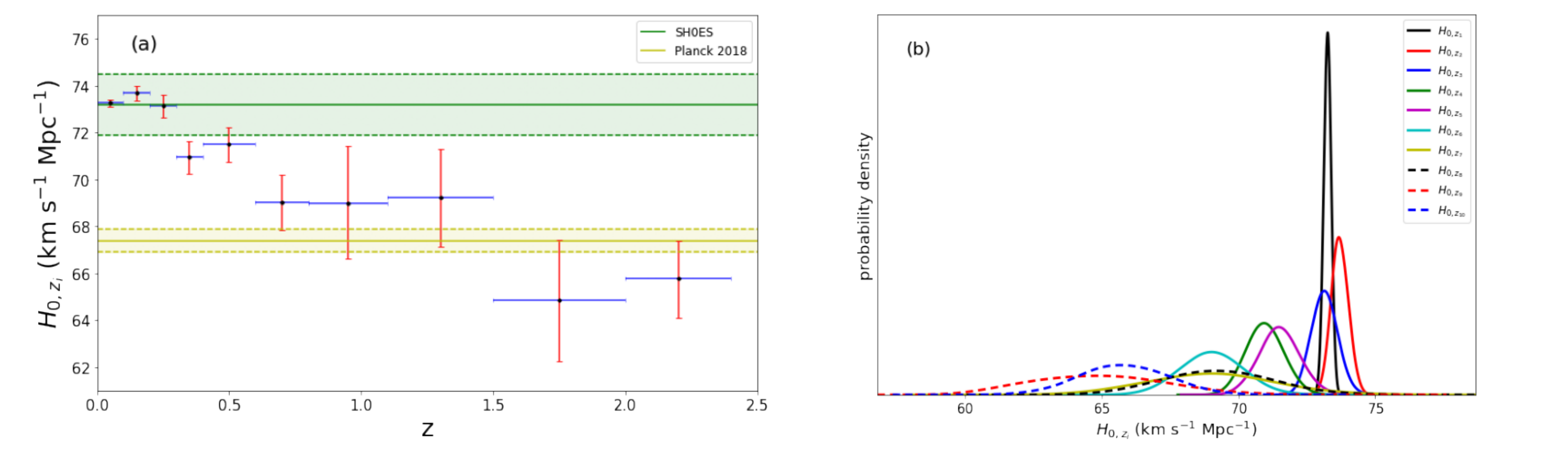}
\vspace{-6pt}
\caption{Fitting 
 results for $H_0(z)$ using equal-width binning across ten redshift intervals. (\textbf{a}): 
 $H_0(z)$ as a function of redshift. A clear decreasing trend is observed, with a significance of $5.6\sigma$ at $z>0.3$. (\textbf{b}): Normalized probability distributions of $H_0(z)$ for ten redshift bins, which are well-approximated by Gaussian profiles. More details can be seen in Ref. \cite{2023A&A...674A..45J}.}
\label{fig:12}
\end{figure}

By combining SNe, BAO, and $H(z)$ data within a Markov Chain Monte Carlo (MCMC) framework, they obtained the posterior distribution of $H_0$. However, the inferred $H_0$ values across different redshift bins are not statistically independent: since distance is a cumulative quantity, this leads to intrinsic correlations among the estimates. To decorrelate these measurements, a transformation matrix was calculated following \mbox{\citet{2005PhRvD..71b3506H}.} Principal Component Analysis (PCA) was then employed to compress the observational constraints while retaining the full information content. The evolution of $H_0$ was investigated using various samples and binning schemes. The results demonstrate that the declining trend in $H_0$ is robust and insensitive to the choice of binning method or sample selection. Specifically, using equal-width binning, a significant $5.6\sigma$ declining trend was detected, as illustrated in Figure~\ref{fig:12}. This data-driven result reveals a clear evolution of $H_0$ with redshift: at low redshifts, the inferred $H_0$ aligns with the local distance ladder \mbox{value \cite{2022ApJ...934L...7R},} while at high redshifts, it converges toward the value inferred from CMB \mbox{measurements \cite{2020A&A...641A...6P}.} This redshift-dependent behavior effectively alleviates the current Hubble tension.

With the availability of additional observational data, \citet{2025ApJ...979L..34J} later conducted a more detailed and refined analysis of this work. The Dark Energy Spectroscopic Instrument (DESI) collaboration has released its cosmological constraints based on baryon acoustic oscillations \cite{DESI_DR1,DESI_DR2}. The results provide hints of dynamic dark energy behavior \cite{2024PhRvD.110l3519J,2025JCAP...09..031C,2025JCAP...10..023L,2026ApJ...999..190C}. When combined with Planck CMB data and SNe Ia, the preference for the $w_0w_a$CDM model over the $\Lambda$CDM model increases to $3.9 \sigma$. The magnitude of this deviation varies depending on the specific SNe sample used \cite{2026MNRAS.548ag632P,2026MNRAS.548ag615O}. This implies that inter-calibration or other systematics also play a role \cite{2025MNRAS.538..875E}.
By incorporating the latest BAO measurements and updated SNe Ia samples, the analysis yielded a consistent declining trend in $H_0$, in agreement with earlier findings \cite{2025ApJ...979L..34J}. The declining trend has been consistently confirmed across different binning schemes and independent observational samples, reinforcing \mbox{its robustness.}

\begin{figure}[H]%
   \hspace{-3mm} \includegraphics[width=0.8\textwidth]{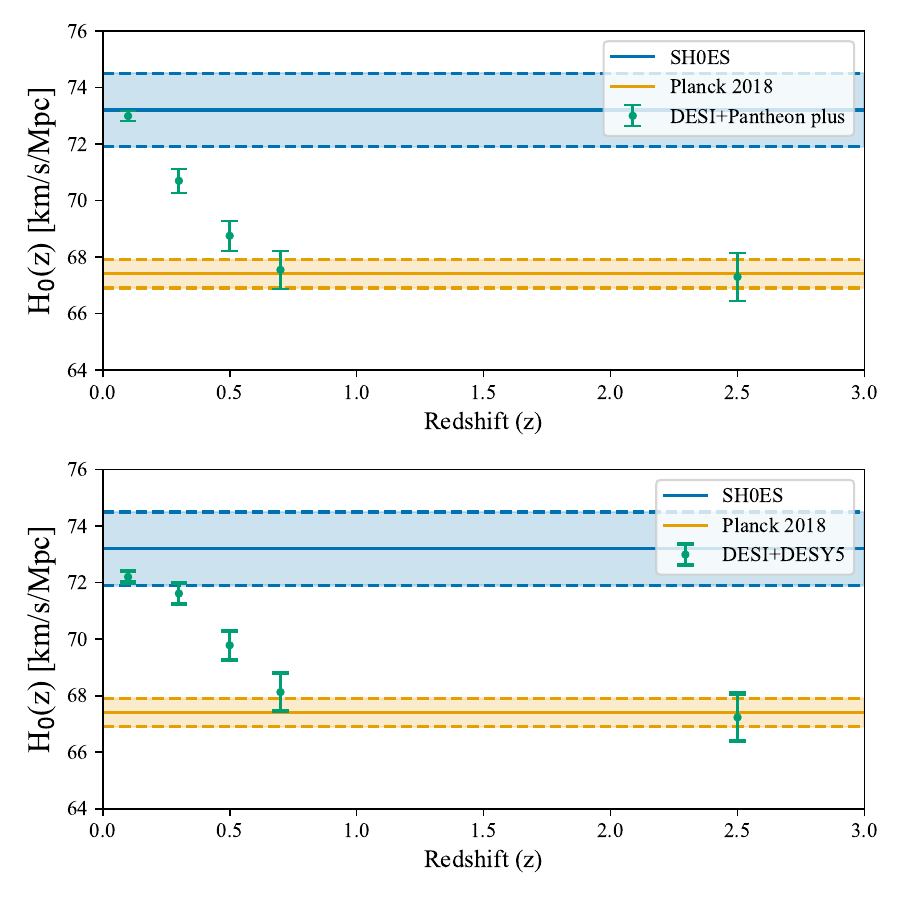} 
   \vspace{-9pt}
    \caption{The 
 descending trend of the Hubble constant $H_0$ derived from the dark energy equation of state $w(z)$. (Top panel). 
 Green data points represent the maximum a posteriori estimates of $H_0(z)$ with $1\sigma$ error bars, obtained from the combination of DESI DR2 BAO measurements and the Pantheon plus SNe Ia sample. The redshifts of these points correspond to the midpoints of their respective bins. The inferred $H_0$ is consistent with the local measurement within $1\sigma$ at low redshift and gradually converges toward the Planck CMB value at high redshift. This redshift dependence effectively mitigates the Hubble tension. (Bottom panel). Same as Top panel, but for the DESI DR2 BAO measurements and DESY5 SNe Ia sample. In both panels, local measurements of $H_0$ have not been used as a constraint. More details can be seen in Ref. \cite{2025ApJ...994L..22J}.}
    \label{fig:13}
\end{figure}

\begin{figure}[H]%
    \includegraphics[width=0.8\textwidth]{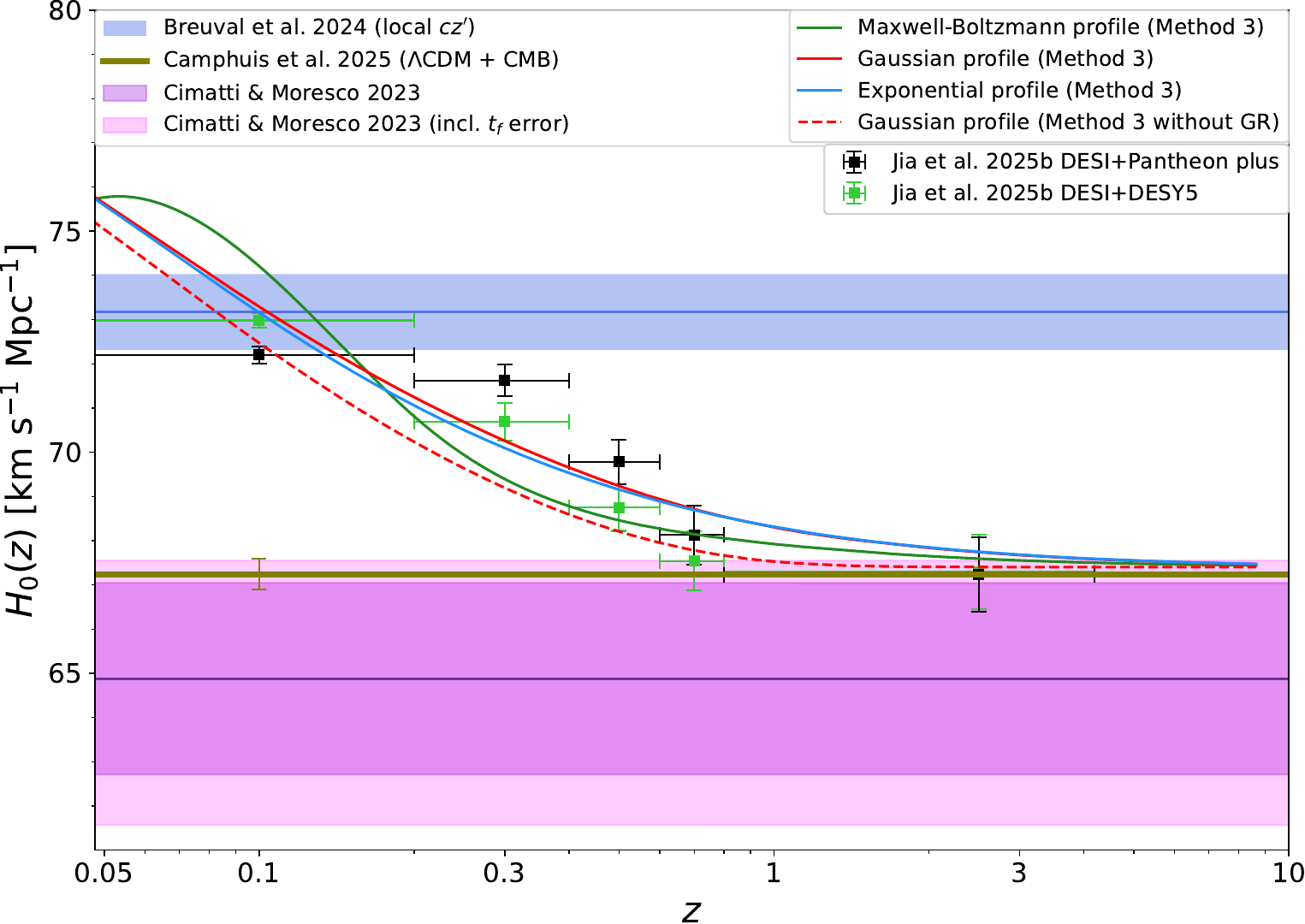} 
    \caption{Predicted 
 $H_0(z)$ curves in the void models and according to observations. The horizontal blue band at the top shows the local $cz'$ measurement \citep{2024ApJ...973...30B}, while the lower magenta band shows $H_0$.}
    \label{fig:14}
\end{figure}

As previously introduced, the currently observed declining trend of $H_0$ is initially inferred directly from data, lacking any corresponding physical explanation. However, as research into this trend has deepened, several theoretical models have been proposed. In a study aimed at interpreting this declining trend, \citet{2025ApJ...994L..22J} explored its connection to the dark energy equation of state. By modeling the evolution of the equation of state, they derived its implied impact on $H_0$ within the $\Lambda$CDM framework. Both SNe and BAO data exhibit deviations from $\Lambda$CDM predictions, with inferred $H_0$ values showing a consistent evolutionary trend, as shown in Figure \ref{fig:13}. Furthermore, \citet{2025MNRAS.536.3232M} proposed using void models to account for this decline. Their results show good agreement with the observed trend under both Gaussian and exponential void density profiles, suggesting that a local void structure could provide a viable physical mechanism to alleviate the Hubble tension while remaining consistent with a low background $H_0$ value, shown in Figure \ref{fig:14}. FRBs hold promise for testing void models \cite{2026arXiv260203928B}. The observed declining trend of $H_0$ provides support for a late-time resolution to the Hubble tension, which could manifest as either a background-level effect or a local void \cite{2026Galax..14...19N}. 

\vspace{-6pt}

{\captionof*{figure}{estimated from old Galactic stars and stellar populations \citep{Cimatti_2023}. The horizontal olive line shows $H_0^\mathrm{CMB}$, with uncertainty indicated at the left \citep{2025arXiv250620707C}. The solid curves show void model \mbox{predictions \citep{2025MNRAS.536.3232M}.} For illustrative purposes, the dashed red curve shows the prediction of the Gaussian model without including gravitational redshift.  
    More details can be seen in Ref. \cite{2025MNRAS.536.3232M}.}}
\vspace{6pt}

\section{Summary and Conclusions} \label{Sec:Summary}
Over the past few decades, cosmology has experienced remarkable progress—driven in large part by the rapid development of robust observational surveys, breakthroughs in instrumental precision, and innovative advances in statistical analysis techniques. In this paper, we provide an overview of the Hubble tension and compile a comprehensive review of various methods used to measure the Hubble constant. Although each measurement method is supported by abundant observational data and well-established analysis techniques, a certain level of inconsistency persists among different datasets interpreted in the standard cosmological model. At present, it appears unlikely that the current Hubble tension can be resolved by any single observational approach alone. We look forward to future joint constraints from multiple observational probes, which hold the promise of improving the current situation \cite{2025arXiv251023823H}. 

Currently, the Hubble tension has surpassed thresholds for statistical significance, reproducibility, and independent cross-validation, and continues to receive sustained attention within the field. Since its emergence, researchers have invested substantial effort into identifying and quantifying potential systematic errors in the underlying measurements. To date, this rigorous scrutiny of systematics has not only improved the reproducibility and robustness of observational results, but has also reinforced—rather than diminished—the evidence for the tension itself. In this paper, we discuss various observational approaches and the systematic uncertainties associated with each. Although some incompletely characterized systematic uncertainties remain in various observations, they are insufficient to account for the observed magnitude of the discrepancy in $H_0$. It may suggest that systematic errors alone are far from sufficient to resolve the current Hubble tension, and that new physics beyond the standard cosmological model may be at play.

Although current local distance ladder measurements have achieved high-precision constraints on the Hubble constant, there remain several avenues for further improvement. The distance ladder is currently supported by geometric distances in four galaxies. Introducing additional anchor points in the calibration of Cepheid variables will reduce uncertainties and strengthen the first rung of the distance ladder. We look forward to upcoming observational advances—such as those from the ongoing Zwicky Transient Facility (ZTF) survey \cite{2017NatAs...1E..71B,2019PASP..131g8001G}, JWST \cite{2023ApJ...956L..18R,2024ApJ...962L..17R}, the next generation of ground-based telescopes (Vera Rubin \mbox{Observatory \cite{2026arXiv260319541F}}) and space telescopes (Nancy Grace Roman \mbox{Observatory \cite{2025AAS...24530509F},} Chinese Space Station Survey Telescope (CSST) \cite{2026SCPMA..6939501C,CSST_zhanhu}), which are expected to provide new insights into the distance ladder. These capabilities will enable deep optical time-domain observations of distant galaxies, leading to the discovery of additional Cepheid variables and significantly expanding the volume of the local universe accessible for Cepheid-based studies. They will also drastically increase the sample of rare astrophysical sources like supernovae, while radio telescopes like the Five-hundred-meter Aperture Spherical Telescope promise similar improvements in FRB detection.

\vspace{6pt}

\authorcontributions{Writing - Original draft, Xuan-Dong Jia, Xin-Yi Dai; Writing - Review \& Editing, Xuan-Dong Jia, Yu-Peng Yang, Fa-Yin Wang.
}

\funding{This work was supported by the National Natural Science Foundation of China (grant Nos. 12494575 and 12273009). 
}

\dataavailability{The relevant data can be found in the corresponding references. 
}

\acknowledgments{We 
 thank the referee for valuable comments and suggestions, which have helped to improve this manuscript. This 
 work was supported by the National Natural Science Foundation of China (grant Nos. 12494575 and 12273009).}

\conflictsofinterest{The authors declare no conflicts of interest. 
}

\begin{adjustwidth}{-\extralength}{0cm}

\reftitle{References}


\PublishersNote{}
\end{adjustwidth}
\end{document}